\documentclass[floatfix,nofootinbib,twocolumn,english,aps,prd,twocolumn,groupedaddress,letterpaper]{revtex4}
\usepackage[T1]{fontenc}
\usepackage[latin1]{inputenc}
\usepackage{color}
\usepackage{graphicx}
\usepackage{bm}

\makeatletter

\providecommand{\tabularnewline}{\\}

\usepackage{array}
\usepackage{hhline}
\usepackage{dcolumn}
\usepackage{rotating}
\usepackage{epstopdf}

\newcommand{\minerva}{Minerva}

\usepackage{babel}
\makeatother
\begin{document}

\preprint{\textbf{\textcolor{blue}{NuSOnG-QCD\_Paper 2}}}

\title{%
QCD Precision Measurements and Structure Function Extraction at a \\
 High Statistics, High Energy Neutrino Scattering Experiment:  NuSOnG \\
}

\date{\today}

\author{T.~Adams$^6$, P.~Batra$^4$, L.~Bugel$^4$,
  L. Camilleri$^4$, J.M.~Conrad$^9$, A.~de~Gouv\^ea$^{12}$,
  P.H.~Fisher$^9$, J.A.~Formaggio$^9$, J.~Jenkins$^{12}$,
  G.~Karagiorgi$^9$, T.R.~Kobilarcik$^5$, S.~Kopp$^{16}$,
  G.~Kyle$^{11}$, W.A.~Loinaz$^1$, D.A.~Mason$^5$, R.~Milner$^9$, R.~Moore$^5$,
  J.~G.~Morf\'{\i}n$^5$, M.~Nakamura$^{10}$, D.~Naples$^{13}$,
  P.~Nienaber$^{14}$, F.I~Olness$^{15}$, J.F.~Owens$^6$,
  S.F. Pate$^{11}$, A. Pronin$^{3}$, W.G.~Seligman$^4$,
  M.H.~Shaevitz$^4$, H.~Schellman$^{12}$, I.~Schienbein$^8$,
  M.J.~Syphers$^5$, T.M.P.~Tait$^{2,12}$, T.~Takeuchi$^{17}$,
  C.Y.~Tan$^5$, R.G.~Van~de~Water$^7$,
  R.K.~Yamamoto$^9$, J.Y.~Yu$^{15}$\\
\bigskip
\bigskip
}

\affiliation{$^1$\/Amherst College, Amherst, MA 01002 \\
$^2$\/Argonne National Laboratory, Argonne , IL 60439 \\
$^{3}$\/Central College, Pella IA 50219\\
$^4$\/Columbia University, New York, NY 10027  \\
$^5$\/Fermi National Accelerator Laboratory, Batavia IL 60510 \\
$^6$\/Florida State University, Tallahassee, FL 32306  \\
$^7$\/Los Alamos National Accelerator Laboratory, Los Alamos, NM 87545  \\
$^8$LPSC, Universit\'{e} Joseph Fourier Grenoble 1, 38026 Grenoble, France\\
$^9$\/Massachusetts Institute of Technology, Cambridge, MA 02139  \\
$^{10}$\/Nagoya University, 464-01, Nagoya, Japan \\
$^{11}$\/New Mexico State University, Las Cruces, NM 88003 \\
$^{12}$\/Northwestern University, Evanston, IL 60208  \\
$^{13}$\/University of Pittsburgh, Pittsburgh, PA 15260  \\
$^{14}$\/Saint Mary's University of Minnesota, Winona, MN 55987\\
$^{15}$\/Southern Methodist University, Dallas, TX 75205 \\
$^{16}$\/University of Texas, Austin TX 78712\\
$^{17}$\/Virginia Tech, Blacksburg VA 24061\\
}

\begin{abstract}
  We extend the physics case for a new high-energy, ultra-high
  statistics neutrino scattering experiment, NuSOnG (Neutrino
  Scattering On Glass) to address a variety of issues including
  precision QCD measurements, extraction of structure functions, and
  the derived Parton Distribution Functions (PDFs).
  This experiment uses a Tevatron-based neutrino beam to obtain a
  sample of Deep Inelastic Scattering (DIS) events which is over two
  orders of magnitude larger than past samples.
  We outline an innovative method for fitting the structure functions
  using a parameterized energy shift which yields reduced systematic
  uncertainties.
  High statistics measurements, in combination with improved
  systematics, will enable NuSOnG to perform discerning tests of
  fundamental Standard Model parameters as we search for deviations
  which may hint of ``Beyond the Standard Model'' physics.
\end{abstract}



\maketitle
\null
\null
\newpage{}
\null
\newpage{}
\tableofcontents{}
\null
\bigskip\bigskip\bigskip\bigskip
\bigskip\bigskip\bigskip\bigskip
\bigskip\bigskip\bigskip\bigskip

\section{Introduction}

\subsection{NuSOnG: Precision Structure Functions and Incisive QCD Measurements\label{QCDinput}}

The search for new physics at the ``Terascale'' -- energy scales of
$\sim$ 1 TeV and beyond -- is the highest priority for particle
physics.

NuSOnG is a proposed high energy, high statistics neutrino scattering
experiment that can search for ``new physics'' from the keV through
TeV energy scales via precision electroweak and QCD measurements.

During its five-year data acquisition period, the NuSOnG experiment
could record almost one hundred thousand neutrino-electron elastic
scatters, and hundreds of millions of Deep Inelastic Scattering (DIS) events,
exceeding the current world data sample by more than an order of
magnitude.

 This experiment can address concerns related to extraction of
structure functions and their derived Parton Distribution Functions (PDFs), investigate nuclear
corrections, constrain isospin violation limits, and perform incisive
measurement of heavy quarks.

\section{Deep Inelastic Scattering and Parton Distribution Functions\label{subsubdis}}
%

\begin{table}[htb]
\begin{tabular}{|c|c|c|c|c|}
\hline 
Experiment &
$\nu$ DIS &
$\bar{\nu}$ DIS &
main &
isoscalar \tabularnewline
&
events &
events &
target &
correction \tabularnewline
\hline 
\hline 
CCFR &
0.95M &
0.17M &
iron &
5.67\% \cite{Seligman:1997fe}\tabularnewline
\hline 
NuTeV &
0.86M &
0.24M &
iron &
5.74\% \cite{shaevitzprivate} \tabularnewline
\hline 
NuSOnG &
606M &
34M &
glass &
isoscalar \tabularnewline
\hline
\end{tabular}
\caption{Comparison of statistics and targets for parton distribution studies
in NuSOnG compared to the two past highest statistics DIS neutrino
scattering experiments.\label{DISstats} }
\end{table}

Obtaining a high quality model of the parton distribution functions
in neutrino and antineutrino scattering is crucial to the NuSOnG electroweak
measurements \cite{Adams:2008cm}. 
NuSOnG will go a step beyond past experiments in addressing
the systematics of parton distribution functions (PDFs) by making
high statistics measurements for neutrino and antineutrino data separately.
Table~\ref{DISstats} shows the large improvement in statistics for
NuSOnG compared to NuTeV and CCFR, the previous highest statistics
experiments. Issues of uncertainties on the nuclear corrections are
avoided by extracting PDFs on SiO$_{2}$ directly, in similar fashion
to the NuTeV Paschos-Wolfenstein (PW)  analysis.

The differential cross sections for neutrino and antineutrino CC DIS
each depend on three structure functions: $F_{2}$, $xF_{3}$ and
$R_{L}$. They are given by: \begin{widetext} \begin{eqnarray}
\frac{d^{2}\sigma^{\nu(\overline{\nu})N}}{dxdy} & = & \frac{G_{F}^{2}ME_{\nu}}{\pi\left(1+Q^{2}/M_{W}^{2}\right)^{2}}\left[F_{2}^{\nu{(\overline{\nu})}N}(x,Q^{2})\left(\frac{y^{2}+(2Mxy/Q)^{2}}{2+2R_{L}^{\nu{(\overline{\nu})}N}(x,Q^{2})}+1-y-\frac{Mxy}{2E_{\nu}}\right)\right.\left.\pm xF_{3}^{\nu{(\overline{\nu})}N}y\left(1-\frac{y}{2}\right)\right],\label{eq:sigsf}\end{eqnarray}
 \end{widetext} where $+(-)$ is for $\nu(\overline{\nu})$ scattering.
In this equation, $x$ is the Bjorken scaling variable, $y$ the inelasticity,
and $Q^{2}$ the squared four-momentum transfer. The structure functions
are directly related to the PDFs.

The function $xF_{3}(x,Q^{2})$ is unique to the DIS cross section
for the weak interaction. It originates from the parity-violating
term in the product of the leptonic and hadronic tensors. For an isoscalar
target, in the quark-parton model, where $s=\bar{s}$ and $c=\bar{c}$,
\begin{eqnarray}
xF_{3}^{\nu N}(x) & = & x\left(u(x)+d(x)+2s(x)\right.\\
 &  & \left.-\bar{u}(x)-\bar{d}(x)-2\bar{c}(x)\right),\nonumber \\
xF_{3}^{\bar{\nu}N}(x) & = & xF_{3}^{\nu N}(x)-4x\left(s(x)-c(x)\right).\label{nubarxF3}\end{eqnarray}
 In past experiments, the average of $xF_{3}$ for neutrinos and antineutrinos
has been measured. Defining $xF_{3}=\frac{1}{2}(xF_{3}^{\nu N}+xF_{3}^{\bar{\nu}N})$,
at leading order in QCD, \begin{equation}
xF_{3,LO}=\sum_{i=u,d..}xq_{i}(x,Q^{2})-x\overline{q_{i}}(x,Q^{2}).\label{DIS xf3 definition}\end{equation}
 To the level that the sea quark distributions have the same $x$
dependence, and thus cancel, $xF_{3}$ can be thought of as probing
the valence quark distributions. The difference between the neutrino
and antineutrino parity violating structure functions, $\Delta(xF_{3})=xF_{3}^{\nu N}-xF_{3}^{\bar{\nu}N}$,
probes the strange and charm seas. ({\it Cf.} Sec.~\ref{CVS}.)

The function $F_{2}(x,Q^{2})$ appears in both the cross section for
charged lepton ($e$ or $\mu$) DIS and the cross section for $\nu$
DIS. At leading order, \begin{equation}
F_{2,LO}=\sum_{i=u,d..}e_{i}^{2}(xq_{i}(x,Q^{2})+x\overline{q_{i}}(x,Q^{2})),\label{DIS F_2 definition}\end{equation}
 where $e_{i}$ is the charge associated with the interaction. In
the weak interaction, this charge is unity. For charged-lepton scattering
mediated by a virtual photon, $e_{i}$ is the fractional electromagnetic
charge of the quark flavor. Thus $F_{2}^{\nu N}$ and $F_{2}^{e(\mu)N}$
are analogous but not identical and comparison yields useful information
about specific parton distribution flavors \cite{Arneodo:1996qe} and charge
symmetry violation as discussed below. In past neutrino experiments,
$F_{2}^{\nu}$ and $F_{2}^{\bar{\nu}}$ have been taken to be identical
and an average $F_{2}$ has been extracted, although this is not necessarily
true in nuclear targets, as discussed below.

Similarly, $R_{L}(x,Q^{2})$, the longitudinal to transverse virtual
boson absorption cross-section ratio, appears in both the charged-lepton
and neutrino scattering cross sections. To extract $R_{L}$ from the
cross section, one must bin in  the variables 
$x$, $Q^{2}$ and $y$. This requires
a very large data set. To date, the best measurements for $R_{L}$  come
from charged lepton scattering rather than 
neutrino scattering \cite{Yang:2001xc}.
Therefore, neutrino experiments have used charged lepton fits to $R_{L}$
as an input to the measurements of $xF_{3}$ and $F_{2}$ \cite{Yang:2000ju}.
This, however, is just a matter of the statistics needed for a global
fit to all of the unknown structure functions in $x$ and $Q^{2}$
bins \cite{Yang:1998qh}. With the high statistics of NuSOnG, precise
measurement of $R_{L}$ will be possible from neutrino scattering
for the first time.

As an improvement on past experiments, the high statistics of NuSOnG
allows measurement of up to six structure functions: $F_{2}^{\nu}$,
$F_{2}^{\bar{\nu}}$, $xF_{3}^{\nu}$, $xF_{3}^{\bar{\nu}}$, $R_{L}^{\nu}$
and $R_{L}^{\bar{\nu}}$. This is done by fitting the neutrino and
antineutrino data separately in $x$, $y$ and $Q^{2}$ as described
in Eq.~(\ref{eq:sigsf}). The first steps toward fitting all six
structure functions independently were made by the CCFR experiment
\cite{McNulty:1997wv}, however statistics were such that only $xF_{3}^{\nu}$,
$xF_{3}^{\bar{\nu}}$, and $F_{2}$-average and $R$-average could
be measured, where the average is over $\nu$ and $\bar{\nu}$. A
global fit of up to  six structure functions in NuSOnG would allow separate
parameterizations of the underlying PDFs which can account for the
nuclear and isospin violation issues discussed below.

In addition to fitting to the inclusive DIS sample, neutrino scattering
can also probe parton distributions through exclusive samples. A unique
and important case is the measurement of the strange sea through 
charged current (CC)
opposite
sign dimuon production. When the neutrino interacts with an $s$ or
$d$ quark, it can produce a charm quark that fragments into a charmed
hadron. The charmed hadron's semi-leptonic decay (with branching ratio
$B_{c}\sim10\%$) produces a second muon of opposite sign from the
first: \begin{eqnarray}
\nu_{\mu}\;+\;{\rm N}\;\longrightarrow\;\mu^{-}\!\! & + & \! c\;+\;{\rm X} 
\nonumber\\
 &  & \!\!\hookrightarrow s\;+\;\mu^{+}\;+\;\nu_{\mu}~.\end{eqnarray}
 Similarly, with antineutrinos, the interaction is with an $\overline{s}$
or $\overline{d}$, \begin{eqnarray}
\overline{\nu}_{\mu}\;+\;{\rm N}\;\longrightarrow\;\mu^{+}\!\! & + & \!\overline{c}\;+\;{\rm X}\nonumber \\
 &  & \!\!\hookrightarrow\!\overline{s}+\;\mu^{-}\;+\;\overline{\nu}_{\mu}~.\end{eqnarray}
 The opposite sign of the two muons can be determined for those events
where both muons reach the toroid spectrometer. Study of these events
as a function of the kinematic variables allows extraction of the
strange sea, the charm quark mass, the charmed particle branching
ratio ($B_{c}$), and the Cabibbo-Kobayashi-Maskaka matrix element,
$|V_{cd}|$.

\def\genfrac#1#2#3#4#5#6{\frac{#5}{#6}}

\section{Experimental Extraction of Structure Functions in NuSOnG}

\subsection{Description of NuSOnG \label{sec:DetectorDescription}}

The NuSOnG detector was designed to be sensitive to a wide range of
neutrino interactions from $\nu $-electron scattering as well as $\nu
$-nucleon scattering, though this paper focuses mainly on the latter
process.  The design has been described in detail in Refs.~\cite{EOI,
Adams:2008cm} and so we provide only a brief summary here.

The neutrino beam would be produced via a high energy external proton
beam from a high intensity accelerator with energy of $\sim 1$ TeV.
The Fermilab Tevatron is an existing example.  The CERN SPS+
\cite{Garoby:2008zz,Garoby:2007at}, which is presently under consideration because of its
value to the LHC energy and luminosity upgrades and to a future beta
beam, is a second example.

\begin{figure}[t]
 \includegraphics[angle=0,width=0.48\textwidth,keepaspectratio]{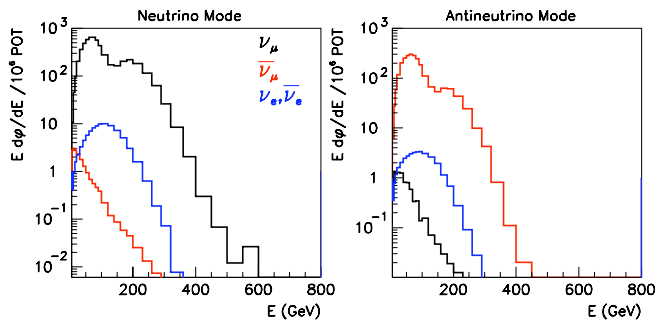}
\caption{The assumed energy-weighted flux
($E\, d\phi/dE/10^6\, POT$)
based on the NuTeV experiment in 
a)~neutrino mode (left) and 
b)~antineutrino mode (right). 
a)~In neutrino mode the fluxes are ordered:
upper (black),  muon neutrino;
middle (blue), electron neutrino and antineutrino;
lower (red),  muon antineutrino.
b)~In antineutrino mode the fluxes are ordered:
upper (red),  muon antineutrino;
middle (blue), electron neutrino and antineutrino;
lower (black),  muon neutrino.
\label{fig:figMike1}}
\end{figure}

The NuSOnG beam design will be based on the one used by the NuTeV
experiment, which is the most recent high energy, high statistics
neutrino experiment. The experiment would use 800 GeV protons on
target followed by a quad-focused, sign-selected magnetic
beam-line. The beam flux, shown in Fig.~\ref{fig:figMike1}, has very
high neutrino or antineutrino purity ($\sim $98\%) and small $\nu
_{e}$ contamination ($\sim $2\%) from kaon and muon decay.  Using an
upgraded Tevatron beam extraction it is expected that NuSOnG could
collect $5\times 10^{19}$ protons/yr, an increase by a factor
of 20 from NuTeV. With this high intensity, such a new facility would
also produce a neutrino beam from the proton dump having a
sizable fraction of tau neutrinos for study.

\begin{figure*}[t]
\includegraphics[angle=-90, width=0.90\textwidth]{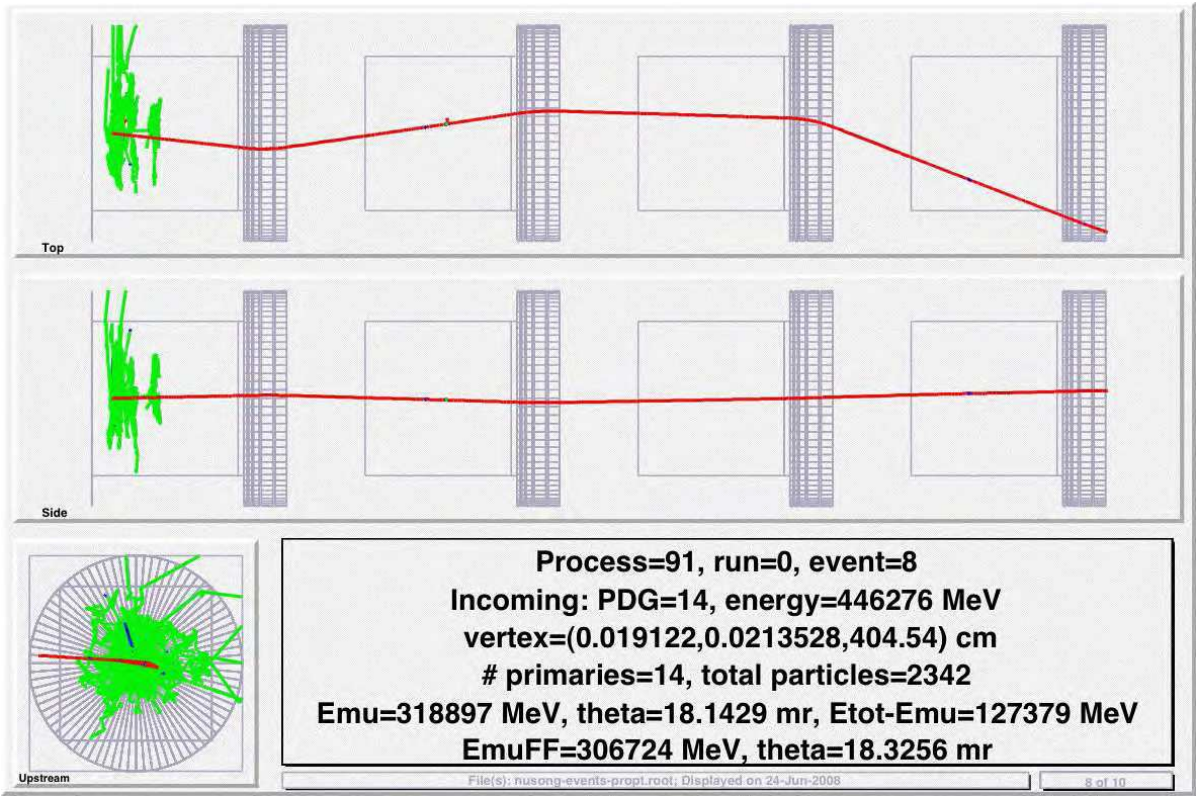}
\caption{A simulated muon neutrino,
charged current event in the NuSOnG detector.
\label{fig:figMike2}}
\end{figure*}

The
baseline detector design is composed of a fine-grained target
calorimeter for electromagnetic and hadronic shower reconstruction
followed by a toroid muon spectrometer to measure outgoing muon
momenta. The target calorimeter will be composed of 2,500 2.5 cm
$\times $ 5 m $\times $ 5 m glass planes interspersed with
proportional tubes or scintillator planes. This gives a target which
is made of isoscalar material with fine 1/4 radiation length
sampling. The detector will be composed of four target sections each
followed by muon spectrometer sections and low mass decay regions to
search for long-lived heavy neutral particles produced in the
beam. The total length of the detector is $\sim $200 m and the
fiducial mass for the four target calorimeter modules will be 3 kiloton
which is 6 times larger than NuTeV or CHARM II. Figure~\ref{fig:figMike2}
shows a simulated $\nu _{\mu }$ charged current event in the detector.

\subsection{Description of NuSOnG Calibration Beam \label{sec:beam}}

The requirements for NuSOnG calibration beam would be similar to
those of NuTeV. Tagged beams of hadrons, electrons, and muons over
a wide energy range (5-200~GeV) would be required.
The calibration beam will have the ability to be steered over the transverse
face of the detector in order to map the magnetic field of each toroid
with muons. Steering for hadrons and electrons
would be less crucial than it was in NuTeV's case, but would still be useful.

The calibration beam can be constructed with a similar design to
NuTeV.
Upstream elements were used to select hadrons, electrons, or
muons. An enhanced beam of electrons was produced by introducing a
thin lead radiator into the beam and detuning the portion of the beam
downstream of the radiator.  A radiator was also used in the nominal
beam tune to remove electrons.  Particle ID (a threshold Cerenkov and
TRDs) was incorporated in the spectrometer and used to tag electrons
when running at low energy.  A pure muon beam was produced by
introducing a 7~m long beryllium filter in the beam as an absorber.

The NuTeV calibration spectrometer determined incoming particle
momenta with a precision of better than 0.3\% absolute  \cite{Harris:1999yg}.  
The NuSOnG goal for calibration-beam precision would be to measure
energy scales to a precision of about 0.5\%, and we demonstrate (in later text
of this paper) that this can be improved with fits to neutrino data.

For comparison, using the calibration beam, NuTeV achieved 0.43\%
precision on absolute hadronic energy scale and 0.7\% on absolute muon
energy scale (dominated by the ability to accurately determine the
toroid map).  Precise knowledge of the muon energy scale is especially
important in order to achieve high measurement accuracy on the
neutrino fluxes using the low-$\nu$ method. For example, a 0.5\%
precision on muon energy scale translates into about a 1\% precision
on the flux.  Both energy scales are important for precision structure
function measurements, and were the largest contributions to structure
function measurement uncertainties in NuTeV \cite{Tzanov:2005kr}.


\subsection{Experimental Extraction of Structure Functions in NuSOnG}

\begin{figure*}[t]
\centering
\includegraphics[totalheight=0.6\textheight]{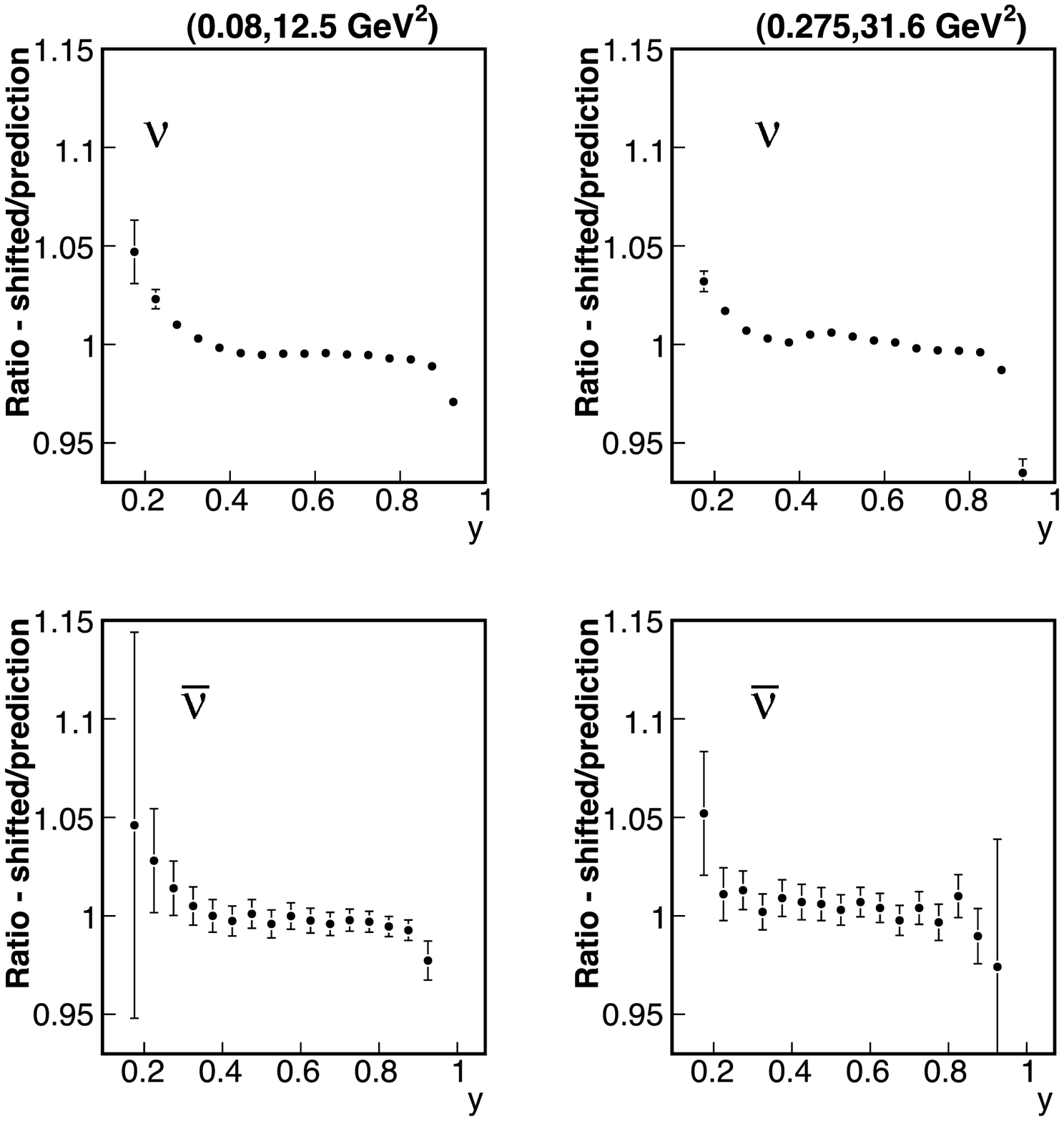}
\caption{Fractional change in number of events for two 
characteristic $(x,Q^2)$ bins
as a function of $y$.  The fractional change comes from scaling the
energy of each event by a factor of 1.005.
}
\label{fi:shift}
\end{figure*}

\begin{figure*}[t]
\centering
\includegraphics[width=0.95\textwidth]{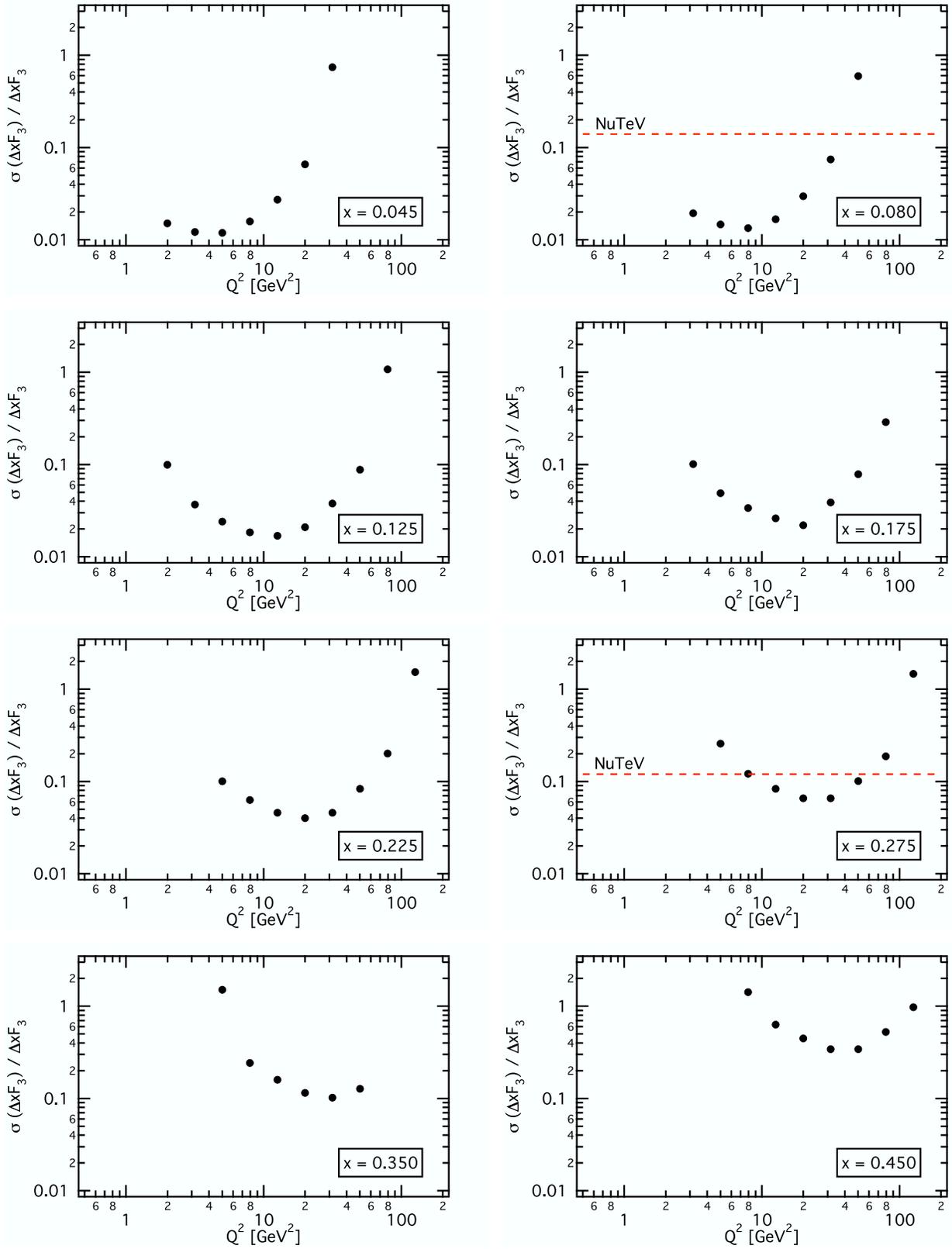}
\caption{Fractional uncertainty for the fit value of $\Delta x F_3$ in
different $x$ bins as a function of $Q^2$.  
The fit is to multiple $x$
and $Q^2$ bins extracting the three structure functions, 
$F_{2}$,$xF_{3}^{avg}$ and $\Delta xF_{3}$.  For each of the fits, a global
set of energy scale parameters is also determined from the fit.
The dotted lines show the
fractional error for the NuTeV 2$\mu$ measurement.}
\label{fi:dxferr}
\end{figure*}

\begin{figure*}[t]
 \includegraphics[angle=0,width=0.90\textwidth,keepaspectratio]{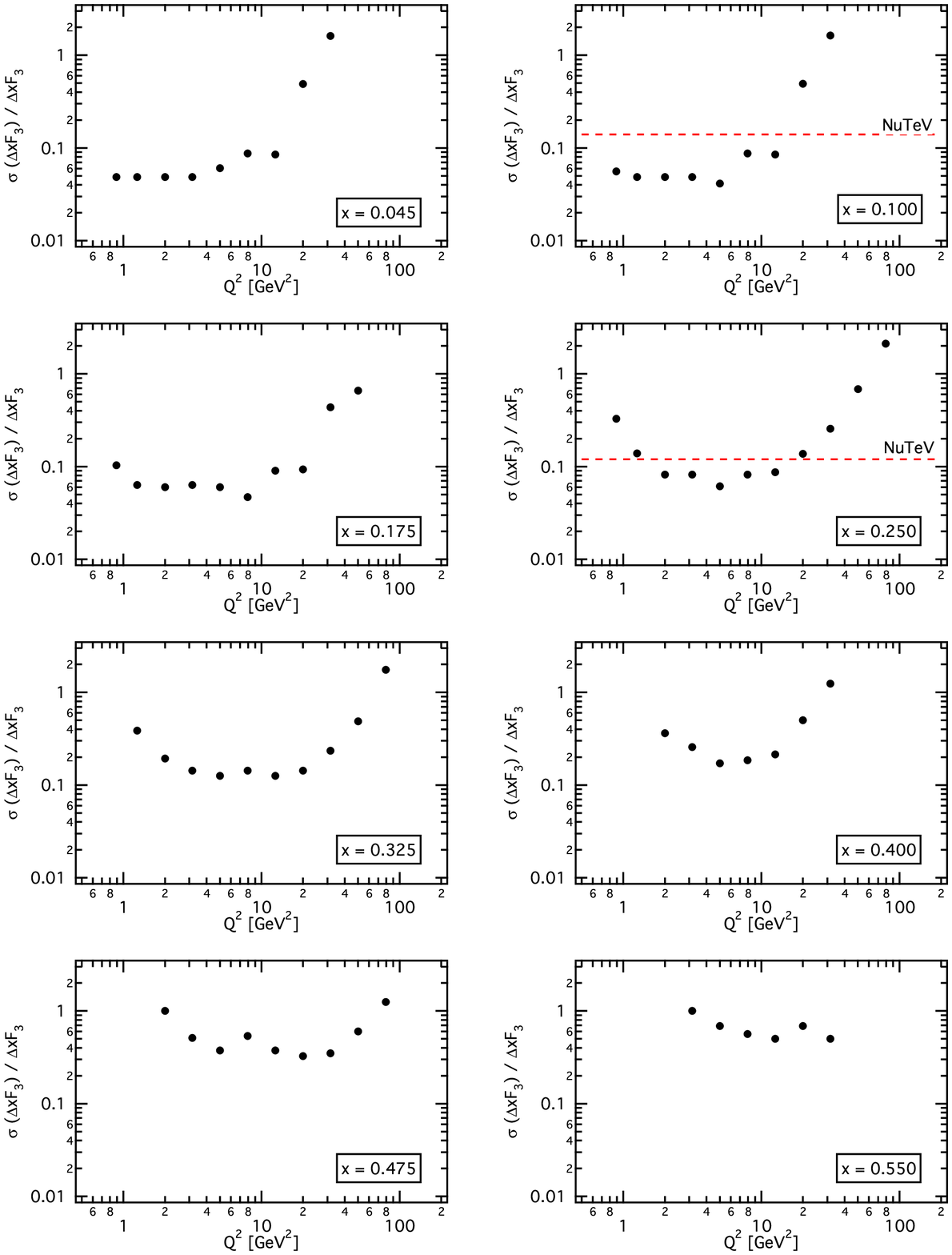}
\caption{Fractional uncertainty for the fit value of $\Delta x F_3$ in
different $x$ bins as a function of $Q^2$. The fit is to multiple $x$
and $Q^2$ bins extracting the four structure functions, 
$F_{2}$,$xF_{3}^{avg}$, $\Delta xF_{3}$, and $R.$  For each of the fits, a global
set of energy scale parameters is also determined.
\label{fi:dxf3err4par}}
\end{figure*}
\begin{figure*}[t]
 \includegraphics[angle=0,width=0.90\textwidth,keepaspectratio]{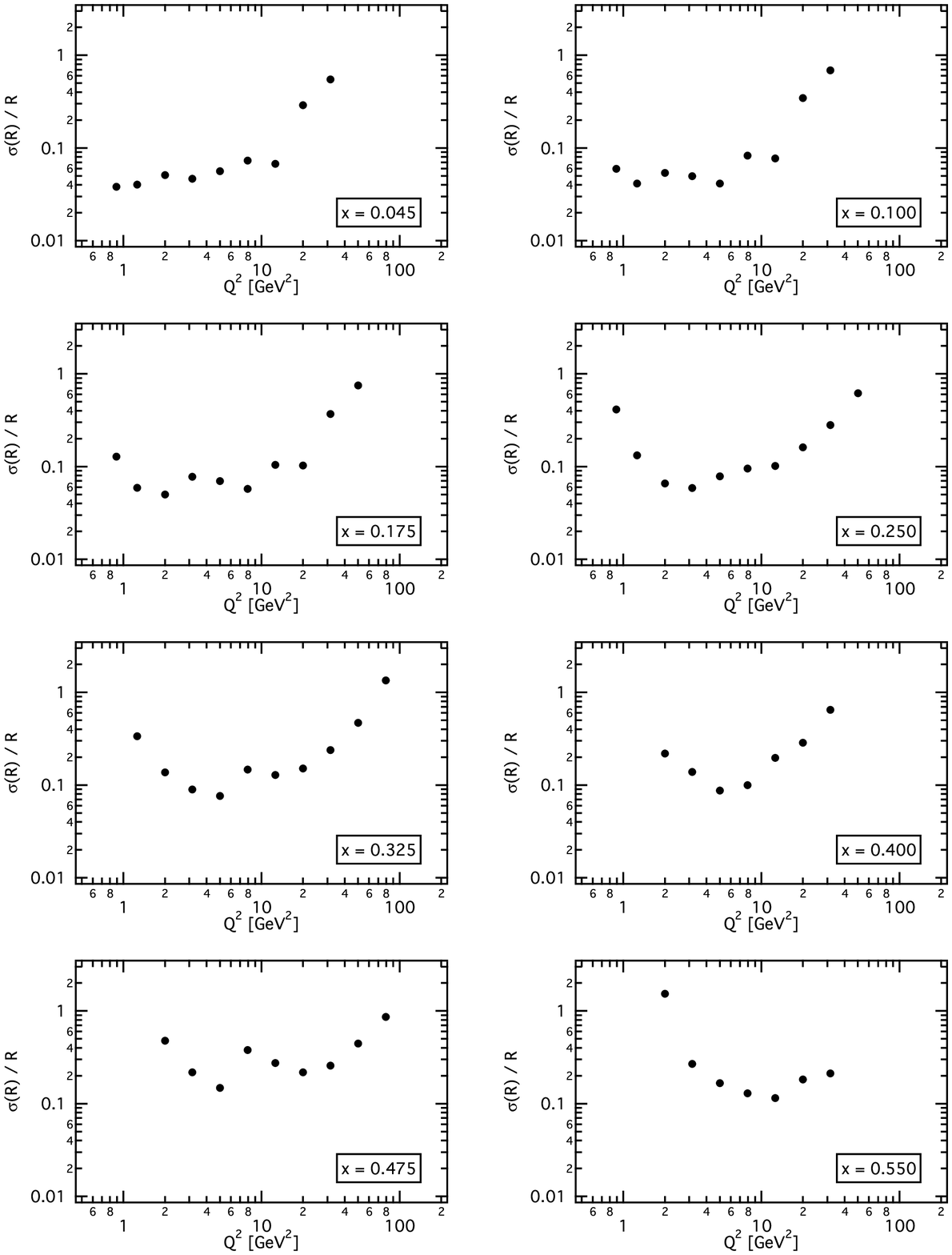}
\caption{Fractional uncertainty for the fit value of $R$ in
different $x$ bins as a function of $Q^2$. The fit is to multiple $x$
and $Q^2$ bins extracting the four structure functions, 
$F_{2}$,$xF_{3}^{avg}$, $\Delta xF_{3}$, and $R.$  For each of the fits, a global
set of energy scale parameters is also determined.
\label{fi:rlongerr4par}}
\end{figure*}

\begin{figure*}[t]
 \includegraphics[angle=0,width=0.90\textwidth,keepaspectratio]{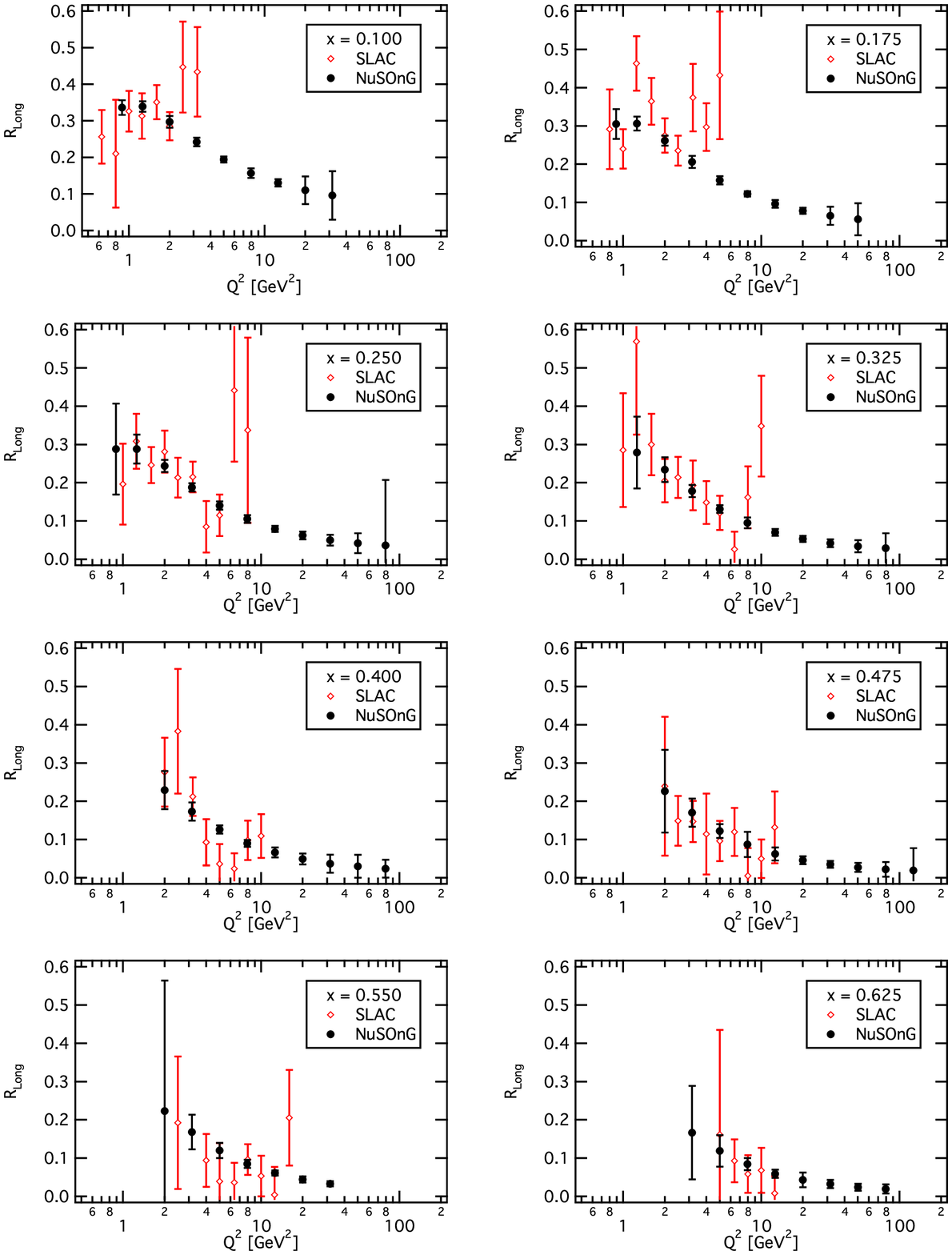}
\caption{Extracted values of $R$ (labeled NuSOnG) from the four structure function 
fits as compared
to previous measurements\cite{Whitlow:1990gk} (labeled SLAC).
\label{fi:rlong4par}}
\end{figure*}

The high statistics of the NuSOnG experiment makes it possible to
extract the structure functions directly from the $y$-distributions
within bins of $\left( {x,Q^{2} } \right) $. Previous lower-statistics
high-energy neutrino experiments either extracted structure functions
by comparing the number of $\nu$ versus $\overline{\nu}$ events in an
$\left( {x,Q^{2} } \right) $ bin \cite{Seligman:1997fe}, or by
extracting the cross-sections ${{d\sigma} \mathord{\left/ {\vphantom
{{d\sigma } {dy}}} \right.  \kern-\nulldelimiterspace} {dy}} $ within
the $\left( {x,Q^{2} } \right) $ bin and fitting for the structure
functions using Equation~(\ref{eq:sigsf}) \cite{Tzanov:2005kr}. Either
method assumes a value for $R_{L}=\sigma _{L}/\sigma_{T}$ as measured
by other experiments \cite{Whitlow:1990gk}, and depends on a
measurement of the strange sea from dimuon events
\cite{Bazarko:1994tt,Tzanov:2003gq}. With sufficient statistics, we
can explore the possibility of measuring $xF_{3 }^{\nu}\left( {x,Q^{2}
} \right) $, $xF_{3 }^{\overline\nu} \left( {x,Q^{2} } \right) $,
$F_{2} \left( {x,Q^{2} } \right) $, and $R \left( {x,Q^{2} } \right) $
from the same data \cite{McNulty:1997wv}.

Let us denote Eq.~(\ref{eq:sigsf}) as a function of the structure
functions by 
$d\sigma^{\nu\left( {\overline{\nu}}\right) }
\left( {xF_{3},F_{2} ,R}\right) $, where the 
$\left( {x,Q^{2}}\right)$-dependence 
is assumed and where the structure functions can be
different for neutrinos and antineutrinos. A sample of Monte Carlo
events, ${N_{{MC,gen}}^{\nu\left( {\overline{\nu}}\right) }}$, is
generated using an assumed set of structure functions for the
cross-section: $d\sigma^{\nu,\overline{\nu}}\left(
{xF_{3}^{gen},F_{2}^{gen},R^{gen}}\right) $. One can then fit for the
structure functions in each $\left( {x,Q^{2}}\right) $ bin by
minimizing
\begin{equation}
\chi^{2}=\sum\limits_{\nu,\overline{\nu}}{\sum\limits_{y-\mathrm{{bins}}%
}{\frac{\left( {N_{data}^{\nu\left( {\overline{\nu}}\right) }%
-N_{MC,pred}^{\nu\left( {\overline{\nu}}\right) }\left(
{SF}_{fit}\right) }\right) ^{2}}{{N_{data}^{\nu\left(
{\overline{\nu}}\right) }}}} },
\label{eq:sfchisq}
\end{equation}
where $N_{MC,pred}^{\nu\left(  {\overline{\nu}}\right)  }\left(  {SF}%
_{fit}\right)  $, the reweighted Monte-Carlo events in an $\left(  {x,Q^{2}%
,y}\right)  $ bin, is given by
 \begin{widetext}\begin{equation}
N_{MC,pred}^{\nu\left(  {\overline{\nu}}\right)  }\left(  {SF}_{fit}\right)
=\sum\limits_{%
\genfrac{}{}{0pt}{}{\scriptstyle\nu\left(  {\overline{\nu}}\right)
\;\mathrm{{events}}\;\mathrm{{in}}\;\hfill}{\scriptstyle\left(  {x,y,Q^{2}%
}\right)  \;\mathrm{{bin}}\hfill}%
}{\frac{{d\sigma^{\nu\left(  {\overline{\nu}}\right)  }\left(  {xF_{3}%
^{fit},F_{2}^{fit},R^{fit}}\right)  }}{{d\sigma^{\nu\left(  {\overline{\nu}%
}\right)  }\left(  {xF_{3}^{gen},F_{2}^{gen},R^{gen}}\right)  }}}%
N_{MC,gen}^{\nu\left(  {\overline{\nu}}\right)  }\left(  {SF}_{gen}\right),
\label{eq:mctheory}%
\end{equation} \end{widetext}
$N_{data}^{\nu\left( {\overline{\nu}}\right) }$ is the number of
${\nu}$ or ${\overline{\nu}}$ data events in the $\left(
{x,y,Q^{2}}\right) $ bin, and ${N_{{MC,gen}}^{\nu\left(
{\overline{\nu}}\right) }}$ is the number of Monte-Carlo events
generated in the $\left( {x,y,Q^{2}}\right) $ bin.
$xF_{3}^{fit}\left( {x,Q^{2}}\right) $, $F_{2}^{fit}\left( {x,Q^{2}%
}\right) $, and $R^{fit}\left( {x,Q^{2}}\right) $ are the fit
parameters in the $\chi^{2}$-minimization of
Eq.~(\ref{eq:sfchisq}). In principle they can be fit separately for
$\nu$ and $\overline{\nu}$ structure functions. Here we will concentrate on the measurement ox
f up to four separate structure
functions, $\Delta xF_{3}\left( {x,Q^{2}}\right) =xF_{3}^{\nu}\left(
{x,Q^{2}}\right) -xF_{3}^{\overline{\nu}}\left( {x,Q^{2}}\right) $,
$xF_{3}^{avg}=(xF_{3}^{\nu}+xF_{3}^{\overline{\nu}})/2$, $F_{2}\left(
{x,Q^{2}}\right) $, and $R\left( {x,Q^{2}}\right) $ where we assume
that $F_{2}\left( {x,Q^{2}}\right) $ and $R\left( {x,Q^{2}}\right) $
are the same for neutrinos and antineutrinos i.e. 
 $F_{2}\left(  {x,Q^{2}}\right)
=F_{2}^{\nu}\left(  {x,Q^{2}}\right)  =F_{2}^{\overline{\nu}}\left(  {x,Q^{2}%
}\right)  $ and $R\left(  {x,Q^{2}}\right)  =R^{\nu}\left(  {x,Q^{2}}\right)
=R^{\overline{\nu}}\left(  {x,Q^{2}}\right)  $.


\subsection{Fitting for $\Delta x F_{3}$}


We have studied the extraction of the structure function from the 600 million
neutrino and 33 million anti-neutrino deep inelastic scattering events
expected in the full NuSOnG data set. The dominant systematic error comes from
the measurement of the muon momentum in the toroidal spectrometer. At NuTeV,
the systematic uncertainty was 0.7\% and we assume NuSOnG will achieve 0.5\%.
Our studies are carried out by fitting the $y$-distribution in each $x,Q^{2}$
bin for $F_{2},$ the average value of $xF_{3}=xF_{3}^{avg}=(xF_{3}^{\nu
}+xF_{3}^{\overline{\nu}})/2$, $\Delta xF_{3}=xF_{3}^{\nu}-xF_{3}%
^{\overline{\nu}},$and $R$. In the first set of studies, $R(x,Q^{2})$ is set
equal to the measured value\cite{Whitlow:1990gk} and fits are done to the
three structure functions, $F_{2}$, $xF_{3}^{avg}$, and $\Delta xF_{3}$.

Our fitting procedure begins with a sample of Monte Carlo generated events,
$N^{gen}(x,Q^{2},y)$, sampled from the CCFR structure functions and the
nominal value for $\Delta xF_{3}$ from NuTeV. We fit in bins of $(x,Q^{2})$ as
a function of $y$ and obtain the fit spectra by reweighting the original
sample: \begin{widetext}
\[
N^{fit}(x,Q^2,y)=\frac{F_2^{fit}(x,Q^2)(2-2y+y^2/(1+R))\pm x
F_3^{fit}(x,Q^2)(1-(1-y)^2)}{F_2^{nom}(x,Q^2)(2-2y+y^2/(1+R))\pm x
F_3^{nom}(x,Q^2)(1-(1-y)^2)}N^{gen}(x,Q^2,y).
\]
\end{widetext}where the upper sign is for neutrinos and the lower for
anti-neutrinos. In order to study the effects of the systematic energy scale
shift, we produce a Monte Carlo sample where the muon energy scale is shifted
by 0.5\%, $E_{\mu}^{meas}=1.005E_{\mu}^{true},$ for each event. The fractional
change in the number of events in each bin due to the energy scale shift is
shown in Fig.~\ref{fi:shift}. 

This shifted event distribution, $N^{shift}(x,Q^{2},y),$ is then used to carry
out a three parameter fit to Eq. \ref{eq:sfchisq} where $F_{2}$, $xF_{3}%
^{avg}$, and $\Delta xF_{3}$ are varied. Large shifts in $\Delta xF_{3}$
result. For example, the shift from the input value in the $(x,Q^{2}%
)=(0.08,12.6GeV^{2})$ bin is 19.01\% and the shift in other bins is even larger.

The effects of the energy scale uncertainty can be practically eliminated by
including energy scale shift parameters in the fit. A muon energy scale change
shifts the events in the various y-bins by an amount that is not consistent
with that expected from changes in the structure functions. Therefore, fits to
the y-distributions can isolate the effects of an energy scale shift and
significantly reduce the structure function uncertainty from this systematic
error. To estimate the systematic error reduction for this technique, three
additional energy scale parameters are introduced in the fit to the
y-distributions. These three parameters are used to produce an energy scale
shift parameterization in each $\left(  x,Q^{2},y\right)  $ bin given by
\[
E_{\mu scale}=E_{\mu scale1}+E_{\mu scale2}E_{\mu}+E_{\mu scale3}E_{\mu}^{2}.
\]
The updated prediction for the number of events in a given $\left(
x,Q^{2},y\right)  $ bin is
\begin{eqnarray}
&&N_{pred}^{\nu\left(  {\overline{\nu}}\right)  }(SF_{fit})=N_{MC,pred}
^{\nu\left(  {\overline{\nu}}\right)  }\left(  {SF}_{fit}\right)  
\nonumber \\
&&\quad 
+E_{\mu
scale}\left(  N^{shift}(x,Q^{2},y)-N^{gen}(x,Q^{2},y)\right) ,
\nonumber \\
\quad  .
\end{eqnarray}
and the $\chi^{2}$ used in the minimization similar to Eq. \ref{eq:sfchisq}
with the addition of pull terms associated with the three energy scale
parameters
\begin{eqnarray}
\chi^2&=&\sum\limits_{\nu,\overline{\nu}}{\sum\limits_{y-\mathrm{{bins}}%
}{\frac{\left(  {N_{data}^{\nu\left(  {\overline{\nu}}\right)  }-N_{pred}%
^{\nu\left(  {\overline{\nu}}\right)  }\left(  {SF}_{fit}\right)  }\right)
^{2}}{{N_{data}^{\nu\left(  {\overline{\nu}}\right)  }}}}}
\nonumber \\
&+&E_{\mu
scale1}+\frac{E_{\mu scale2}}{(0.02)^{2}}+\frac{E_{\mu scale3}}{(0.0002)^{2}%
}.\label{eq:sfchisqpull}%
\end{eqnarray}
These pull terms correspond to an energy scale uncertainties of about 0.5\%
for muon energy values averaging between 50 and 70 GeV. This fitting technique
renders the systematic error from the scale shift to be small in comparison
with the statistical error. For example, in the bin $\left(  x,Q^{2}\right)
=$ (0.275, 32 GeV$^{2}$) bin, the systematic error for $\Delta xF_{3}$ is
0.3\% while the statistical error is 10\%; the value of the $E_{\mu scale1}$
parameter is also determined to about 10\%.

In the ultimate analysis, the fit will be carried out simultaneously over all
$x$ and $Q^{2}$ bins with one set of energy scale parameters. We have studied
this using eight $x$ bins and six to eight $Q^{2}$ bins.
Figure~\ref{fi:dxferr} shows the fractional error on 
$\Delta xF_{3}$ for different $x$ bins as a
function of $Q^{2}$. In general, we believe NuSOnG can measure $\Delta xF_{3}$
over most of the $(x,Q^{2})$ range to better than 10\%; in many cases around 3\%. 
Typical
values for NuTeV are shown in two $x$ bins in Fig.~\ref{fi:dxferr}. Since more
than one $\left(  x,Q^{2}\right)  $ bin is being used to determine the energy
scale shift parameters, the value of the $E_{\mu scale1}$
parameter can also determined to about 3\% from these fits.

Simulation studies have also been made to estimate the uncertainties
associated with doing fits to extract the four structure functions, $F_{2},$
$xF_{3}^{avg}$, $\Delta xF_{3},$and $R.$ The procedure is the same as used for
the three structure function fits where the $\chi^{2}$ in Eq.
\ref{eq:sfchisqpull} is minimized simultaneously over a number of $x$ and
$Q^{2}$ bins with one set of energy scale parameters. In this case, the
$\Delta xF_{3}$ and $R_{long}$ structure functions can be determined to
between 5\% and 20\% for most of the $x$ and $Q^{2}$ range as shown in Figs.
\ref{fi:dxf3err4par} and \ref{fi:rlongerr4par}. The simulated $R_{long}$
measurements are shown in Fig. \ref{fi:rlong4par} along with previous
measurements.\cite{Whitlow:1990gk} As indicated from this figure, the
capabilities of the NuSOnG to measure $R_{long}$ is much more precise that any
previous experiment.

In summary, due to the very high statistics of a NuSOnG type experiment, an
almost complete set of structure functions over a broad range of $x$ and
$Q^{2}$ can be extracted from the data without introducing theoretical or
experimental approximations. Further, systematic uncertainties that have
limited the precision of previous structure function measurements can be
eliminated by including fits to these uncertainties in the extraction
procedure. We believe that with these techniques the structure function
measurements will be statistics limited even for NuSOnG.

\section{Nuclear Effects \label{sec:nukes}}

Historically, neutrino experiments have played a major role in expanding
our understanding of parton distribution functions through high statistics
experiments such as CCFR~\cite{Yang:2000ju}, 
NuTeV~\cite{Tzanov:2005kr,Fleming:2000bg,Yang:2000ju},
and CHORUS~\cite{Onengut:2005kv}. However, the high statistics extract
a price since the large event samples require the use of nuclear targets
-- iron in the case of both CCFR and NuTeV and lead in the case of
the Chorus experiment. The problem is that if one wants to extract
information on \textit{nucleon} PDFs, then the effects of the nuclear
targets must first be removed. NuSOnG can provide key measurements
which will improve these corrections.

Charged lepton deep inelastic scattering has been measured on a wide range of 
targets. The most simplistic expectation for the structure functions might be 
that they would simply be given by an average of the appropriate number of 
proton and neutron results as in 

$$F_2^A(x,Q^2) = \frac{Z}{A} F_2^P(x,Q^2) + \frac{A-Z}{A} F_2^n(x,Q^2).$$ 

\begin{figure}[t]
\includegraphics[angle=-90, width=0.48\textwidth,keepaspectratio]{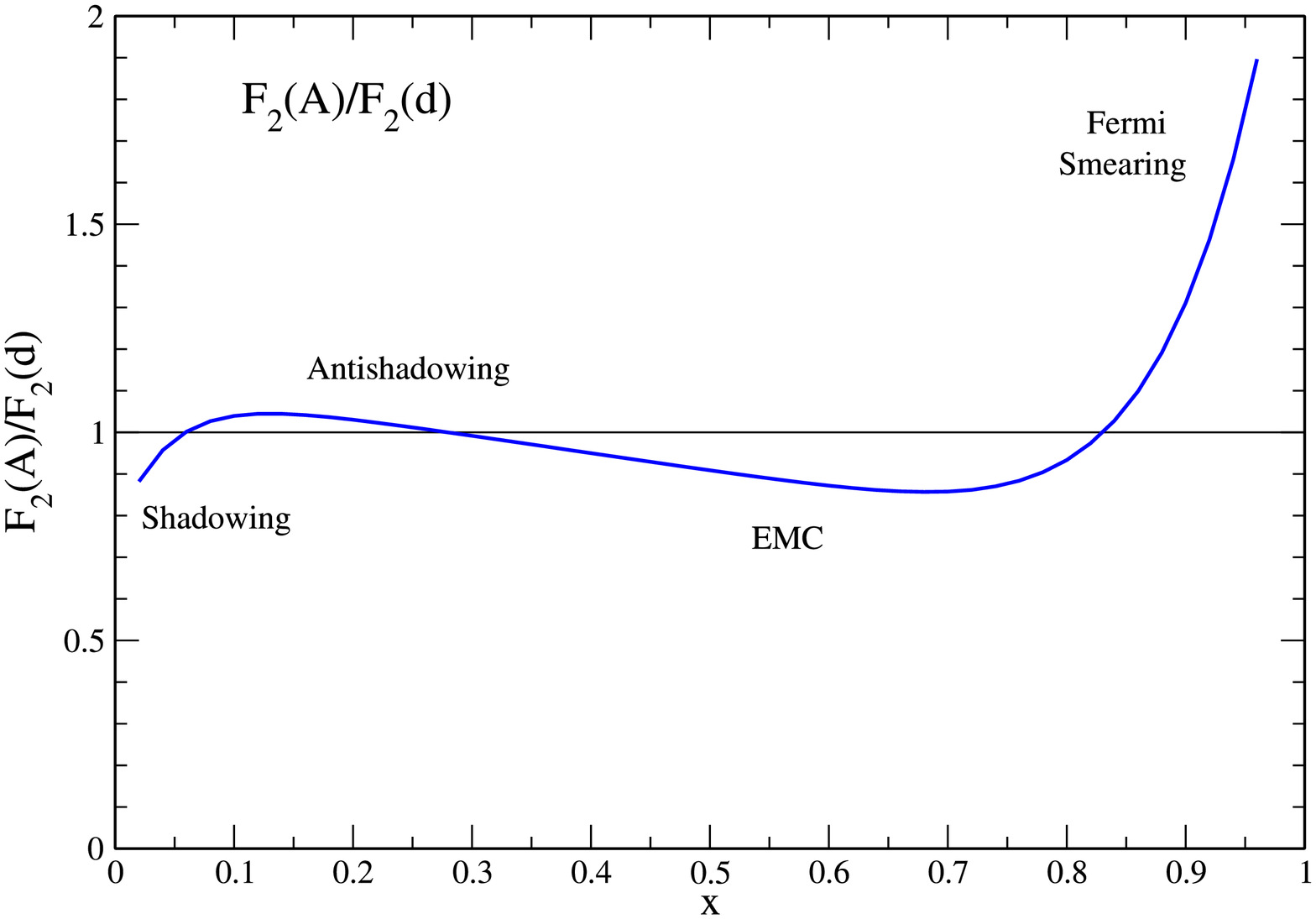}
\caption{Typical behavior of the ratio of the structure function on a nuclear 
target $A$ to that on a deuterium target.}
\label{emc_ratio}
\end{figure}

However, the results from a wide range of experiments show a much more complex 
behavior for the structure functions on nuclei. The typical behavior of 
the ratio of $F_2^A(x,Q^2)$ to $F_2^d(x,Q^2)$ where $d$ denotes a deuterium 
target shows four distinct regions as sketched in Fig.~\ref{emc_ratio}.

At small $x$ the ratio dips below one in what is called the shadowing region. 
At somewhat larger values of $x$ the ratio rises above one in the 
antishadowing region. At still larger values of $x$ the ratio again falls below 
one in the EMC region. Finally, as $x$ approaches one, Fermi motion smearing 
causes a significant rise in the ratio.

This behavior shows only a modest dependence on $A$ for values above beryllium, 
with the shape remaining qualitatively the same and the amount of the 
suppression at $x\approx 0.6$ increasing slowly with $\log(A)$. Furthermore, 
there is little, if any, observed dependence on $Q^2$. These features are 
summarized nicely in the results shown in Ref.~\cite{Gomez:1993ri}.

The mechanisms of nuclear scattering have also been studied theoretically.
These mechanisms appear to be different for small and large Bjorken
$x$ as viewed from the laboratory system. Bjorken $x$ is defined
as $x=Q^{2}/2M\nu$, where $\nu$ and $\bm{q}$ are energy and momentum
transfer to the target and $Q^{2}=\bm{q}^{2}-\nu^{2}$. The physical
quantity which is responsible for the separation between large and
small $x$ regions is a characteristic scattering time, which is also
known as Ioffe time (or length) $\tau_{I}=\nu/Q^{2}$~\cite{Ioffe:1985ep}.
If $\tau_{I}$ is smaller than the average distance between bound
nucleons in a nucleus then the process can be viewed as incoherent
scattering off bound nucleons. This happens at larger $x(>0.2)$.

\subsection{Nuclear effects at small $x$}

\label{sec:smallx}

We expect to find a difference between charged-lepton nucleus and
neutrino nucleus scattering at small-$x$ because the space-time pictures
for the two processes are different in this region. The underlying physical 
mechanism in
the laboratory reference frame can be sketched as a two-stage process.
At the first stage, the virtual photon $\gamma^{*}$, or $W^{*}$
or $Z^{*}$ in case of neutrino interactions, fluctuates into a quark-gluon
(or hadronic) state. In the second stage, this hadronic state then
interacts with the target. The uncertainty principle allows an estimate
of the average lifetime of such hadronic fluctuation as \begin{equation}
\tau=\frac{2\nu}{m^{2}+Q^{2}}=\frac{1}{x\, M}\,\frac{Q^{2}}{m^{2}+Q^{2}},
\label{tau}\end{equation}
where $m$ is invariant mass of hadrons into which the virtual boson
convert, and $M$ is the proton mass. The same scale $\tau$ also
determines characteristic longitudinal distances involved in the process.
At small $x$, $\tau$ exceeds the average distance between bound
nucleons. For this reason coherent multiple interactions of this hadronic
fluctuation in a nucleus are important in the small-$x$ kinematical
region. It is well known that the nuclear shadowing effect for structure
functions is a result of coherent nuclear interactions of hadronic
fluctuations of virtual intermediate boson.
\footnote{For a recent review of nuclear shadowing see, e.g., 
\cite{Piller:1999wx}.}

For neutrino interactions which are mediated by the axial-vector current,
the fluctuation time $\tau$ is also given by Eq.~\ref{tau}. However,
as was argued in Ref.~\cite{Kopeliovich:1992ym}, the fluctuation and coherence
lengths are not the same in this case. In particular, the coherence
length is determined by the pion mass $m_{\pi}$ in Eq.~\ref{tau} 
because of the dominance of off-diagonal transitions like $a_{1}N\to\pi N$
in nuclear interactions. Since the pion mass is much smaller than
typical masses of intermediate hadronic states for the vector current
($m_{\rho},\ m_{\omega}$, etc.), the coherence length $L_{c}$ of
intermediate states of the axial current at low $Q^{2}$ will be much
larger than $L_{c}$ of the vector current. A direct consequence of
this observation is the early onset of nuclear shadowing in neutrino
scattering at low energy and and low $Q^{2}$ as compared with the
shadowing in charged-lepton scattering. The basic reason for this
earlier onset and different behavior in the transition region is the
difference in the correlation lengths of hadronic fluctuations of
the vector and axial-vector currents. This is also illustrated by
observing that for a given $Q^{2}$, the cross-section suppression
due to shadowing occurs for much lower energy transfer ($\nu$) in
neutrino interactions than for charged leptons.

\subsection{Nuclear Effects in Neutrino Interactions}

As there has been no systematic experimental study of 
$\nu\ \mbox{and}\ \overline{\nu}$
nucleus interactions, one must then rely on theoretical models of
the nuclear corrections. This is an unsatisfactory situation since
one is essentially measuring quantities sensitive to the convolution
of the the desired PDFs and unknown -- or model dependent -- nuclear
corrections.

As noted above, theoretically there are substantial differences between
charged lepton and neutrino interactions on the same nucleus. There
are other expected differences for neutrinos. For example, the relative
nuclear shadowing effects for the structure function $F_{3}$ is predicted
to be substantially different from that for $F_{2}$~\cite{Kulagin:1998wc}.
This is because the structure function $F_{3}$ describes the correlation
between the vector and the axial-vector current in neutrino scattering.
In terms of helicity cross sections, the structure function $F_{3}$
is given by the cross section asymmetry between the left- and right-polarized
states of a virtual $W$ boson. It is known that such a difference
of cross sections is strongly affected by Glauber multiple scattering
corrections in nuclei. \cite{Badelek:1992gs,Kulagin:2007ju,Glauber:1970jm}
This causes an enhanced nuclear shadowing effect
for the structure function $F_{3}$.

It is important to experimentally address the question of nuclear
effects in neutrino scattering so that the neutrino data can be used
in proton fits without bringing in substantial nuclear uncertainties.
For example, in a recent analysis~\cite{Owens:2007kp} the impact of new
neutrino data on global fits for PDFs was assessed. The conclusion
reached in this analysis was that the uncertainties associated with
nuclear corrections precluded using the neutrino data to constrain
the nucleon PDFs. If NuSOnG can address these uncertainties, then
the neutrino data can play an even more prominent role in the global
fits to the proton PDF.

Furthermore, nuclear effects are interesting in their own right. 
Parameterizations of nuclear PDFs on various targets exist in the literature. 
However, there is no universally accepted model which describes these nuclear 
corrections over the entire range of $x$ from first principles. This makes it 
difficult to generalize the above behavior observed in charged lepton DIS to 
DIS with $\nu \ \mbox{or}\ \overline{\nu}$ beams. Models such as that in 
Ref.~\cite{Kulagin:2004ie} exist, but to date there have been no high 
statistics studies 
of $\nu \ \mbox{or}\ \overline \nu$ DIS over a wide range of nuclear targets 
with which to test them.

\begin{figure}[t]
\includegraphics[clip,width=0.48\textwidth,keepaspectratio]{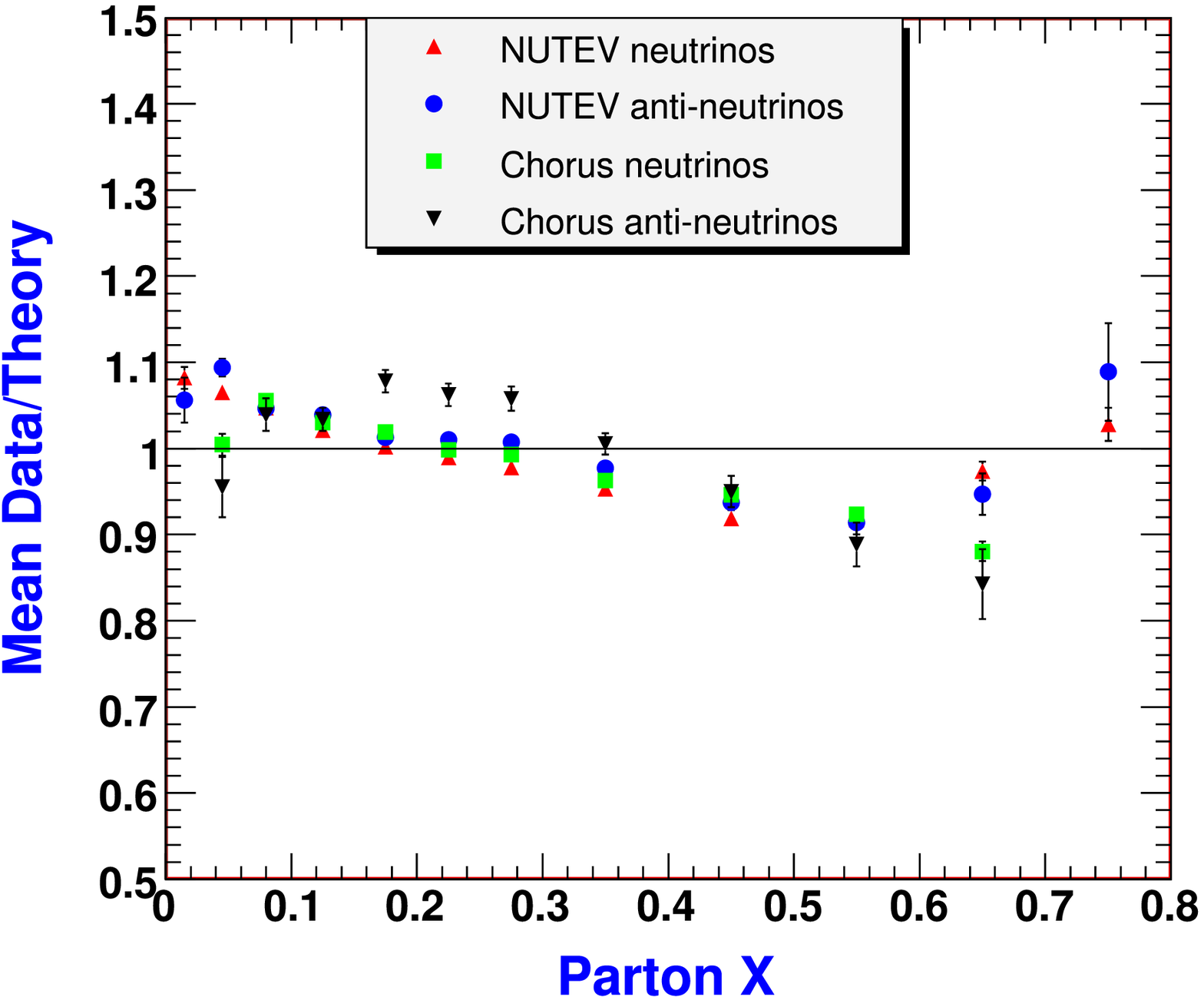} 
\caption{Comparison between the reference fit and the unshifted CHORUS and
NuTeV neutrino data without any nuclear corrections.}
\label{no_shift_no_cor} 
\end{figure}

A study presented in Ref.~\cite{Owens:2007kp} examined the role of new lepton 
pair 
production data from E-866 and new neutrino DIS data from the NuTeV and CHORUS 
collaborations in global fits for nucleon PDFs. For the actual fitting of the 
PDFs it was necessary to include nuclear corrections for the neutrino 
and antineutrino cross sections and the model of Ref.~\cite{Kulagin:2004ie} 
was used. 
As a byproduct of that analysis, it was possible to compare a reference fit, 
obtained without using data on nuclear targets, to the neutrino and 
antineutrino data in order to obtain an estimate of what the nuclear 
corrections should look like. This comparison is shown in 
Fig.~\ref{no_shift_no_cor}.  

This figure shows some results from Ref.~\cite{Owens:2007kp} in the form of 
{}``data/theory''
averaged over $Q^{2}$ and presented versus $x$. The results are
from a global fit but are plotted \textit{without} the model-dependent
nuclear corrections which were used in the fit (the neutrino data were 
\textit{not} used in the reference fit.)
It is notable that the overall pattern of deviations shown in 
Fig.~\ref{no_shift_no_cor}
are, in general, similar to that seen in charged lepton DIS as sketched 
in Fig.\ref{emc_ratio}.
However, the deviations from unity are perhaps smaller. At high $x$,
the effect of Fermi smearing is clear. At moderate $x$ the EMC effect
is observable. It is interesting to note that there is no clear indication
of a turnover at low $x$ in the shadowing region for $\nu$ data. 
Also, note the striking similarity between the $\nu\ \mbox{and}\ \overline{\nu}$
results. This appears to imply that the differences in the nuclear
effects between neutrino and antineutrino DIS are small. As discussed
later, when we consider $\Delta xF_{3}$ and isospin violation, it
is crucial to model differences in the nuclear effects between $\nu$
and $\bar{\nu}$ scattering as a function of $x$. 

To make progress in understanding nuclear corrections in neutrino
interactions, access to high statistics data on a variety of nuclear
targets will be essential. This will allow the $A$-dependence to
be studied as a function of both $x\ \mbox{and}\ Q^{2}$, as has been
done in charged lepton deep inelastic scattering. PDFs from global
fits without the neutrino data can then be used to make predictions
to be compared with the $A$-dependent $\nu\ \mbox{and}\ \overline{\nu}$
cross sections, thereby allowing the nuclear corrections to be mapped
out for comparison with theoretical models.

The primary target of NuSOnG will be SiO$_{2}$. However, we can investigate
a range of $A$-values by replacing a few slabs of glass with alternative
target materials: C, Al, Fe, and Pb. This range of nuclear targets
would both extend the results of \minerva\ to the NuSOnG kinematic
region, and provide a check (via the Fe target) against the NuTeV
measurement.

Given the NuSOnG neutrino flux, we anticipate $58k$ $\nu$-induced
and $30k$ $\bar{\nu}$-induced CC DIS events per ton of material.
A single ton would be sufficient to extract $F_{2}(x)$ and $xF_{3}(x)$
averaged over all $Q^{2}$; a single 5 m$\times$5 m$\times$2.54
cm slab of any of the above materials will weigh more than that. The
use of additional slabs would permit further extraction of the structure
functions into separate $(x,Q^{2})$ bins as was done in the NuTeV
analysis, at the potential expense of complicating the shower energy
resolution in the sub-detectors containing the alternative targets;
this issue will be studied via simulation.

Table~\ref{ta:nuclear} shows that two 50-module stacks would be
sufficient to accumulate enough statistics on alternative nuclear
targets for a full structure-function extraction for each material.
However, for basic cross-section ratios in $x$, a single slab of
each would suffice.

\begin{table}
\centering \begin{tabular}{|c||c||c|}
\hline 
Material  & Mass of  & Number of slabs needed \tabularnewline
 & 2.54 cm slab (tons)  & for NuTeV-equivalent statistics \tabularnewline
\hline
\hline 
C  & 1.6  & 33 \tabularnewline
Al  & 1.9  & 27 \tabularnewline
Fe  & 5.5  & 10 \tabularnewline
Pb  & 7.9  & 7 \tabularnewline
\hline
\end{tabular}

\caption{Alternative target materials for cross-section analysis}

\label{ta:nuclear} 
\end{table}

\subsection{Measuring Nuclear Effects with the \minerva\ and NuSOnG Detectors}

The \minerva\ experiment will also be studying neutrino induced
nuclear effects and will be starting its initial physics run in early
2010. To study nuclear effects in \minerva, a cryogenic vessel containing
liquid helium (0.2 ton fiducial mass)will be installed upstream of
the \minerva\ detector. Within the \minerva\ detector, solid carbon,
iron and lead targets will be installed upstream of the pure scintillator
active detector. The total mass is 0.7~ton of Fe, 0.85~ton of Pb,
0.4~ton of He and somewhat over 0.15 ton of C. Since the pure scintillator
active detector essentially acts as an additional 3-5 ton carbon target
(CH), the pure graphite (C) target is mainly to check for consistency.
For a run consisting of 4.0 x $10^{20}$ POT in the NuMI Low Energy
(LE) beam and $12\times10^{20}$ POT in the NuMI Medium Energy (ME)
beam, \minerva\ would collect over 2~M events on Fe, 2.5~M events
on Pb, 600~K on helium and 430~K events on C as well as 9.0~M events
on the scintillator within the fiducial volume.

Studying nuclear effects with the NuSOnG detector will involve fewer
nuclear targets but considerably more statistics on each. In addition,
the much higher energy of the incoming neutrinos with NuSOnG means
a much wider kinematic range of study. In particular, NuSOnG will
have a much higher Q$^{2}$ for a given low-$x$ to study shadowing
by neutrinos and will be able to measure the shadowing region down
to much smaller $x$ for the same Q$^{2}$ range as \minerva. A significant
addition to the study of nuclear effects with neutrinos would be the
addition of a large, perhaps active (\char`\"{}Bubble Chamber\char`\"{}),
cryogenic target containing hydrogen or deuterium. With the intense
NuSOnG neutrino beam, a significant sample of neutrino-hydrogen and
neutrino-deuterium events could provide the normalization we need
to further unfold nuclear effects in neutrino-nucleus interactions.

\section{QCD Fits}
\newcommand{\overset}{\overline}

The extraction of up to six structure functions from the cross sections
of neutrino and anti-neutrino DIS discussed so far ({\it cf}, Eq.~(\ref{eq:sigsf})) 
has been completely
model-independent relying only on some fundamental principles
such as Lorentz-invariance of the cross section 
and gauge-invariance of the hadronic tensor which is expanded
in terms of the structure functions which parameterize the unknown 
hadronic physics. 

More can be said about the structure functions in QCD. 
While it is still not possible to accurately compute the
$x$-dependence of the structure functions from first principles,
QCD allows us to derive renormalization group equations (RGEs) 
which relate the structure functions at different (perturbative) 
scales $Q$.
Note that the structure functions at the scale $Q$ 
can be directly related to structure functions at a different scale $Q_0$
(see, e.g., Eqs.\ (5.58) and (5.76) in \cite{Reya:1979zk}).
However, it is more convenient to work in the
QCD-improved parton model where the RGEs governing the scale-dependence of
the parton distribution functions (PDFs) are the familiar DGLAP
evolution equations; these can also be used to compute the structure functions at Q given the PDFs at that scale.
\cite{Dokshitzer:1977sg,Gribov:1972ri,Altarelli:1977zs}
Furthermore, this approach has the crucial advantage that the 
universal PDFs allow us to make predictions for other
observables as well.
In addition to the $Q$-dependence, the QCD calculations provide
certain (approximate) relations between different structure functions  
as will be visible from the parton model expressions below.

In this section we will discuss the analysis of the cross section data
within the framework of the QCD-improved parton model. 
Already in the past, high statistics measurements of neutrino
deeply inelastic scattering (DIS) on heavy nuclear targets (NuTeV, ...) 
have attracted much  interest in the literature
since they provide valuable information for global fits of 
PDFs \cite{Martin:2007bv,Nadolsky:2008zw}.

Due to the weak nature of neutrino interactions, the use of nuclear targets is unavoidable; this complicates the extraction of free nucleon PDFs, because model-dependent corrections must be applied to the data ({\it cf} Sec.~\ref{sec:nukes}).
Of course, these same data are also useful for extracting the 
{\em nuclear} parton distribution functions (NPDFs) and 
for such an analysis no nuclear correction factors are required.
Conversely, the NPDFs can be utilized to compute the required 
nuclear correction factors within the QCD parton model \cite{Schienbein:2007fs}.
Similar to proton PDFs, universal nuclear PDFs are needed for 
the description of many processes with nuclei in the initial state.
This involves physics at other neutrino experiments, heavy ion colliders
(RHIC, LHC), and a possible future electron-ion collider (EIC).

The NuSOnG experiment will have two orders of magnitude higher statistics 
than the NuTeV and CCFR experiments
(over an extended kinematic range), and so it will be possible to study small effects such as 
the strangeness asymmetry with better precision, or to establish
for the first time isospin violation in the light quark sector. 
Better understanding these effects is relevant for improving the 
extraction of the weak mixing angle in a Paschos--Wolfenstein type analysis.

\subsection{PDFs}

NuSOnG will perform measurements on different nuclear targets.
The PDFs for a nucleus $(A,Z)$ are constructed as
\begin{equation}
f_i^A(x,Q) = \frac{Z}{A}\  f_i^{p/A}(x,Q) 
+ \frac{(A-Z)}{A}\  f_i^{n/A}(x,Q)\, .
\label{eq:pdf}
\end{equation}

In the following discussion we take into account deviations
from isospin symmetry, a non-vanishing strangeness asymmetry
and the possibility to have non-isoscalar targets.
For this purpose we introduce the following linear combinations
of strange quark PDFs:
\begin{equation}
s^{+,A} = s^A + \bar{s}^A
\, , \quad
s^{-,A} = s^A - \bar{s}^A \, ,
\end{equation} 
where the strangeness asymmetry is described
by a non-vanishing PDF $s^-$.
Note however that  we continue to assume $s^{p/A} = s^{n/A}$ and
$\bar{s}^{p/A} = \bar{s}^{n/A}$.
Also, we neglect any possible charm asymmetry, i.e.,
we use $c^A = \bar{c}^A$ such that $c^{-,A} = c^A - \bar{c}^A = 0$
and $c^{+,A} = c^A + \bar{c}^A = 2 c^A$.

Deviations from isospin symmetry can be parameterized in the following way:
\begin{eqnarray}
\delta u_v^{p/A} &=& u_v^{p/A} - d_v^{n/A}\, , \,
\delta d_v^{p/A} = d_v^{p/A} - u_v^{n/A}\, , \,
\label{eq:delta1}
\\
\delta \bar{u}^{p/A} &=& \bar{u}^{p/A} - \bar{d}^{n/A}\, , \,
\delta \bar{d}^{p/A} = \bar{d}^{p/A} - \bar{u}^{n/A}\, .
\label{eq:delta2}
\end{eqnarray} 
\noindent
These definitions allow us to write the PDFs in a way which makes deviations from 
isoscalarity and isospin symmetry manifest:
\begin{eqnarray}
\label{eq:deviation1}
2 u_v^A &=& [u_v^{p/A} + d_v^{p/A} - \delta d_v^{p/A}] -
\nonumber\\
& & \Delta [u_v^{p/A} - d_v^{p/A} + \delta d_v^{p/A}]\, ,
\\
\label{eq:deviation2}
2 d_v^A &=& [u_v^{p/A} + d_v^{p/A} - \delta u_v^{p/A}] +
\nonumber\\
& & \Delta [u_v^{p/A} - d_v^{p/A} - \delta u_v^{p/A}]\, ,
\\
\label{eq:deviation3}
2 \bar{u}^A &=& [\bar{u}^{p/A} + \bar{d}^{p/A} - \delta \bar{d}^{p/A}] -
\nonumber\\
& & \Delta [\bar{u}^{p/A} - \bar{d}^{p/A} + \delta \bar{d}^{p/A}]\, ,
\\
\label{eq:deviation4}
2 \bar{d}^A &=& [\bar{u}^{p/A} + \bar{d}^{p/A} - \delta \bar{u}^{p/A}] +
\nonumber\\
& & \Delta [\bar{u}^{p/A} - \bar{d}^{p/A} - \delta \bar{u}^{p/A}]\, ,
\end{eqnarray}
where $\Delta = (N-Z)/A$ parameterizes the deviation from isoscalarity.
We have written Eqs.(\ref{eq:deviation1})--(\ref{eq:deviation4}) so that the RHS 
is expressed explicitly in terms of proton PDFs and the four $\delta$-terms 
$\{
\delta u_v^{p/A}, 
\delta d_v^{p/A}, 
\delta \bar{u}^{p/A}, 
\delta \bar{d}^{p/A}
\}$; 
the $\delta$-terms vanish individually if isospin symmetry is preserved.

\subsection{Structure functions}

The structure functions for a nuclear target $(A,Z)$ are given by
\begin{equation}
F_i^A(x,Q) =  \frac{Z}{A}\  F_i^{p/A}(x,Q) + \frac{(A-Z)}{A}\  F_i^{n/A}(x,Q)
\label{eq:sfs}
\end{equation}
such that they can be computed in next-to-leading order 
as convolutions of the nuclear PDFs with the conventional
Wilson coefficients, {\it i.e.}, generically 
\begin{equation}
F_i^A(x,Q) = \sum_k C_{ik} \otimes f_k^{A}\, .
\label{eq:sfs2}
\end{equation}

In order to discuss which information can be extracted from a high 
statistics measurement of neutrino and anti-neutrino DIS cross sections
we briefly review the parton model expressions for the 6 structure
functions.
For simplicity, we first restrict ourselves to leading order, neglect
heavy quark mass effects (as well as the associated production thresholds),
and assume a diagonal CKM matrix.
In our numerical results, these effects are taken into account.

\begin{figure}[t]
\includegraphics[width=0.48\textwidth,keepaspectratio]{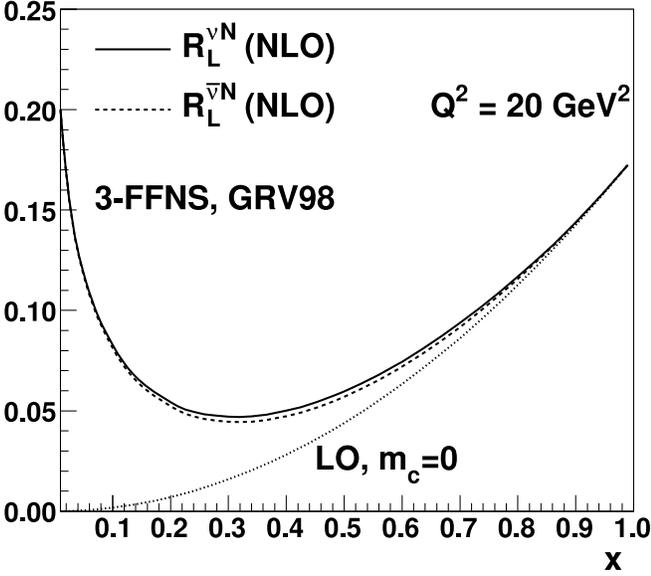} 
\caption{Structure function ratio $R_L$ for neutrino and anti-neutrino nucleon ($N=(p+n)/2$)) 
scattering at $Q^2=20$ GeV$^2$. The solid and dashed lines show NLO results obtained with the 
GRV98 PDFs \protect\cite{Gluck:1998xa}, while 
the dotted line shows the LO result of Eq.\ (\protect\ref{eq:RLLO}).}
\label{fig:RL} 
\end{figure}

The neutrino--nucleus structure
functions are given by (suppressing the dependence on $x$
  and $Q^2$):
\begin{eqnarray}
F_1^{\nu A} &=& d^A + s^A + \bar{u}^A + \bar{c}^A + \ldots \, ,
\\
F_2^{\nu A} &=& 2 x F_1^{\nu A}\, ,
\label{eq:f2nu}
\\
F_3^{\nu A} &=& 2\left [d^A + s^A - \bar{u}^A - \bar{c}^A + \ldots \right] \, .
\end{eqnarray}
The structure functions for anti-neutrino scattering are obtained by exchanging
the quark and anti-quark PDFs in the corresponding neutrino structure functions:
\begin{equation}
F_{1,2}^{\bar{\nu}A} = F_{1,2}^{\nu A}[q \leftrightarrow \bar{q}]\  , \,
\qquad
F_{3}^{\bar{\nu}A} = -F_{3}^{\nu A}[q \leftrightarrow \bar{q}] \  .
\end{equation}
Explicitly this gives
\begin{eqnarray}
F_1^{\bar\nu A} &=& u^A + c^A + \bar{d}^A + \bar{s}^A + \ldots \, ,
\\
F_2^{\bar\nu A} &=& 2 x F_1^{\bar\nu A}\, ,
\label{eq:f2nub}
\\
F_3^{\bar\nu A} &=& 2\left [u^A + c^A - \bar{d}^A - \bar{s}^A + \ldots \right] \, .
\end{eqnarray}
The longitudinal structure function can be obtained with the help
of the following relation:
\begin{equation}
F_L^{\nu A}  = r^2 F_2^{\nu A} - 2 x F_1^{\nu A} 
= \frac{4 x^2 M^2}{Q^2} F_2^{\nu A}
\, ,
\end{equation}
where $r^2 = 1 + 4 x^2 M^2/Q^2$. 
Finally, it is customary to introduce the ratio of longitudinal to transverse
structure functions:
\begin{equation}
R_L^{\nu A} = \frac{F_L^{\nu A}}{2 x F_1^{\nu A}}
= \frac{r^2 F_2^{\nu A}}{2 x F_1^{\nu A}} - 1
= \frac{4 x^2 M^2}{Q^2}\, .
\label{eq:RLLO}
\end{equation}
Similar equations hold for anti-neutrino scattering.
As can be seen, in leading order $R_L^{\nu} = R_L^{\bar{\nu}}$.
As is shown in Fig.\ \ref{fig:RL},
also in NLO, the differences between $R_L^{\nu}$ and $R_L^{\bar{\nu}}$ are tiny 
such that the difference between 
these two functions can be neglected in the following discussion.

\subsection{Constraints on the PDFs}
The differential cross section in Eq.~(\ref{eq:sigsf}) can be
written as:
\begin{equation}
\frac{d \sigma}{dx dy} =  K [A + B (1-y)^2 + C y^2]
\end{equation}
with $K = \frac{G_F^2 M E}{2 \pi (1+Q^2/M_W^2)^2}$,
$A = F_2 \pm x F_3$, $B = F_2 \mp x F_3$, and $C = \frac{2 x^2 M^2}{Q^2} F_2 - F_L$
where the upper sign refers to neutrino and the lower one to anti-neutrino scattering.
This form of $d \sigma$ shows that the (anti-)neutrino cross section data
naturally encodes information on the four structure function combinations
$F_2^{{\nu}A} \pm x F_3^{{\nu}A}$ 
and 
$F_2^{{\bar\nu}A} \pm x F_3^{{\bar\nu}A}$ 
in separate regions of the phase space.
In addition, at large $y$ the structure function combination $C$ contributes.
However, to good accuracy $C^\nu = C^{\bar{\nu}}$ so that $C$ drops out in the
difference of neutrino and anti-neutrino cross sections.

Assuming $s^A=\bar{s}^A$ and $c^A=\bar{c}^A$,
the structure functions $F_2^{{\nu}A}$ and $F_2^{{\bar\nu}A}$ 
constrain the valence distributions
$d_v^A = d^A - \bar{d}^A$, $u_v^A = u^A - \bar{u}^A$ and the flavor-symmetric sea
$\Sigma^A := \bar{u}^A + \bar{d}^A + \bar{s}^A + \bar{c}^A + \ldots$ via the relations:
\begin{eqnarray}
\frac{1}{x} F_2^{\nu A} &=& 2 \left[d_v^A + \Sigma^A \right]\, , 
\\
\frac{1}{x} F_2^{\bar\nu A} &=& 2 \left[u_v^A + \Sigma^A \right]\, .
\end{eqnarray}
Furthermore, we have
\begin{eqnarray}
\frac{1}{x} F_2^{\nu A}+ F_3^{\nu A}&=& 4 (d^A + s^A)\, , 
\\
\frac{1}{x} F_2^{\bar\nu A}- F_3^{\bar\nu A}&=& 4 (\bar{d}^A + \bar{s}^A)\, .
\end{eqnarray}
Since we constrain the strange distribution utilizing the dimuon data,
the latter two structure functions are useful to separately extract the 
$d^A$ and $\bar{d}^A$ distributions.

For an isoscalar nucleus we encounter further simplifications. In this case,
$u^A=d^A$ and $\bar{u}^A = \bar{d}^A =:\bar{q}^A$ which implies $u_v^A = d_v^A =: v^A$.
Hence, the independent quark distributions are 
$\{ v^A, \bar{q}^A, s^A=\bar{s}^A,  c^A=\bar{c}^A,\,  \ldots \}$.
In particular, we have $F_2^{\nu A} = F_2^{\bar{\nu} A}$ for an isoscalar target 
such that our original set of 6 independent structure functions
reduces to 3 independent functions (say $F_2^{\nu A}$, $F_3^{\nu A}$, $F_3^{\bar{\nu} A}$)
under the approximations made.

In a more refined analysis, allowing for a non-vanishing strangeness asymmetry
and isospin violation we can evaluate
the non-singlet structure function
$\Delta F_2^\nu  \equiv  F_2^{\nu A} - F_2^{\bar{\nu}A}$ with the help of the
relations in Eqs.\ (\ref{eq:deviation1}) -- (\ref{eq:deviation4}):
\begin{eqnarray}
\Delta F_2^\nu 
&=& 
2 x s^{-,A} + x\ \delta d_v^{p/A} - x\ \delta u_v^{p/A}
\nonumber\\
&& + \Delta x [2 u_v^{p/A} -2 d_v^{p/A} + \delta d_v^{p/A} - \delta u_v^{p/A}]\, .
\end{eqnarray}
For a nuclear isoscalar target ($Z = N = A/2$, $\Delta = 0$) this expression
simplifies to 
\begin{equation}
\Delta F_2^\nu 
= 2 x s^{-,A} + x\ \delta d_v^{p/A} - x\ \delta u_v^{p/A}\, .
\end{equation}
As one can see, $\Delta F_2^\nu$ will be small and sensitive to
the strangeness asymmetry and isospin violating terms for the valence quarks.

The difference of the neutrino and anti-neutrino cross-sections provides,
in principle, access to this quantity:
\begin{eqnarray}
\frac{d^2 \sigma^{\nu A}}{dx dy} - \frac{d^2 \sigma^{\bar{\nu} A}}{dx dy} 
&\simeq&  K [\Delta F_2^\nu + x F_3^\nu 
\nonumber\\
&&+ (1-y)^2 (\Delta F_2^\nu - x F_3^\nu)]
\end{eqnarray}
with $F_3^\nu = F_3^{\nu A} + F_3^{\bar{\nu}A}$.

It should be noted, however, that in a global fit 
to extract structure functions
 we do not
make direct use of these equations [the $(1-y)^2$-dependence] but simply 
perform a $\chi^2$-analysis of all neutrino and anti-neutrino cross section data.

\section{Isospin (Charge Symmetry) Violation and $\Delta xF_{3}$ \label{CVS}}

The question of isospin violation is central to the PW electroweak
measurement. In the NuTeV analysis, isospin symmetry was assumed.
As discussed in Ref.~\cite{Adams:2008cm}, 
various models which admit isospin violation can
pull the NuTeV $\sin\theta_W$ measurement
 toward the Standard Model. However it would
take significantly larger isospin violation to bring NuTeV into agreement
with the rest of the world's data. Better constraints of isospin violation will be crucial
to the interpretation of the NuSOnG results.

When we relate DIS measurements from heavy targets such as ${}_{26}^{56}{\rm Fe}$
(used in NuTeV) or ${}_{82}^{207}{\rm Pb}$ (Chorus) back to a proton
or isoscalar target, we generally make use of isospin symmetry where
we assume that the proton and neutron PDFs can be related via a $u\leftrightarrow d$
interchange. While isospin symmetry is elegant and well motivated,
the validity of this exact charge symmetry must ultimately be established
by experimental measurement. There have been a number of studies investigating
isospin symmetry violation \cite{Boros:1999fy,Ball:2000qd,Kretzer:2001mb,Martin:2001es,Boros:1998es,Boros:1998qh,Baldit:1994jk};
therefore, it is important to be aware of the magnitude of potential
violations of isospin symmetry and the consequences on the extracted
PDF components. For example, the naive parton model relations are
modified if we have a violation of exact $p\leftrightarrow n$ isospin-symmetry,
or charge symmetry violation (CSV); \textit{e.g.}, $u^{n}(x)\not\equiv d^{p}(x)$
and $u^{p}(x)\not\equiv d^{n}(x)$.

It is noteworthy that a violation of isospin symmetry is automatically
generated once QED effects are taken into account \cite{Martin:2004dh,Roth:2004ti,Gluck:2005xh}.
This is because the photon couples to the up quark distribution $u^{p}(x)$
differently than to the down quark distribution $d^{n}(x)$. These
terms can be as much as a few percent in the medium $x$ range, see
\emph{e.g.} Fig.~1 in Ref.~\cite{Gluck:2005xh}.

Combinations of structure functions can be particularly sensitive
to isospin violations, and NuSOnG is well suited to measure some of
these observables. For example, residual $u,d$-contributions to $\Delta xF_{3}=xF_{3}^{\nu}-xF_{3}^{\bar{\nu}}$
from charge symmetry violation would be amplified due to enhanced
valence components $\{ u_{v}(x),d_{v}(x)\}$, and because the $d\rightarrow u$
transitions are not subject to slow-rescaling corrections which suppress
the $s\rightarrow c$ contribution to $\Delta xF_{3}$ \cite{Kretzer:2001mb}.
Here the ability of NuSOnG to separately measure $xF_{3}^{\nu}$ and
$xF_{3}^{\bar{\nu}}$ over a broad kinematic range will provide powerful
constraints on the sensitive structure function combination $\Delta xF_{3}$.

Separately, the measurement of $\Delta F_{2}\equiv\frac{5}{18}\, F_{2}^{CC}(x,Q^{2})-F_{2}^{NC}(x,Q^{2})$
in Charged Current (CC) $W^{\pm}$ exchange and Neutral Current (NC)
$\gamma/Z$ exchange processes can also constrain CSV \cite{Boros:1998es};
because NuSOnG will measure $F_{2}^{CC}$ on a variety of targets,
this will reduce the systematics associated with the heavy nuclear
target corrections thus providing an additional avenue to study CSV.

In the following, we provide a detailed analysis of CSV which also
investigates the various experimental systematics associated with
each measurement. We shall find it is important to consider all the
systematics which impact the various experimental measurements to
assess the discriminating power.

\subsection{$\Delta xF_{3}$ and Isospin Violations}

We recall the leading-order relations of the neutrino structure function
$F_{3}$ on a general nuclear target:\begin{eqnarray}
\frac{1}{2}F_{3}^{\nu A}(x) & = & d^{A}+s^{A}-\bar{u}^{A}-\bar{c}^{A}+...,\label{eq:f3nu}\\
\frac{1}{2}F_{3}^{\bar{\nu}A}(x) & = & u^{A}+c^{A}-\bar{d}^{A}-\bar{s}^{A}+...\label{eq:f3nubar}\end{eqnarray}
 where $A$ represents the nuclear target $A=\{ p,n,d,...\}$, and
the {}``...'' represent higher-order contributions and terms from
the third generation $\{ b,t\}$ quarks. Note that to illustrate the
general features of these processes, we use a schematic notation as
in Eq.~(\ref{eq:f3nu}) and Eq.~(\ref{eq:f3nubar}); for the numerical
calculations, the full NLO expressions are employed including mass
thresholds, {}``slow-rescaling'' variables, target mass corrections,
and CKM elements where appropriate.

For a nuclear target $A$ we can construct $\Delta xF_{3}^{A}$ as:
\begin{eqnarray}
\Delta xF_{3}^{A} & = & xF_{3}^{\nu A}-xF_{3}^{\bar{\nu}A}\nonumber \\
 & = & 2x\Delta \left[\left(u^{p/A}-d^{p/A}\right)
+\left(\bar{u}^{p/A}-\bar{d}^{p/A}\right) + \frac{1}{2} \delta I^{A} \right]\nonumber \\
 & + & 2x\, s^{+,A}-2xc^{+,A}+x\, \delta I^{A}
+{\cal O}\left(\alpha_{S}\right)\label{eq:deltaf3a}
\end{eqnarray}
 where ${\cal O}\left(\alpha_{S}\right)$ represents the higher order
QCD corrections, and the isospin violations are given by $\delta I^{A}$:
\begin{equation}
\delta I^{A} =  \delta d - \delta u + \delta \bar{d} - \delta \bar{u}\, .
\label{eq:csv}
\end{equation}
 For a flux-weighted linear combination of $F_{3}^{\nu}$ and $F_{3}^{\bar{\nu}}$,
terms proportional to the strange quark asymmetry can enter Eq.~(\ref{eq:deltaf3a}),
\emph{cf.} Refs.~\cite{Boros:1998es,Boros:1998qh,Boros:1999fy}.
For a sign-selected $\nu/\bar{\nu}$ beam as for NuTeV or NuSOnG,
this complication is not necessary. We have defined 
$s^{\pm,A}(x)=[s^{A}(x)\pm\bar{s}^{A}(x)]$
and $c^{\pm,A}(x)=[c^{A}(x)\pm\bar{c}^{A}(x)]$.

In the limit of isospin symmetry, all four terms on the RHS of Eq.~(\ref{eq:csv})
vanish individually. For a nuclear isoscalar target, $Z=N=A/2$, we
can construct $\Delta xF_{3}$ from the above:
\begin{equation}
\Delta xF_{3}=xF_{3}^{\nu A}-xF_{3}^{\bar{\nu}A}=
2xs^{+,A}-2xc^{+,A}+x\,\delta I^{A}+{\cal O}\left(\alpha_{S}\right)
\,.\label{eq:deltaf3}
\end{equation}
 Note in Eq.~(\ref{eq:deltaf3a}) that for a nuclear target $A$
which is close to isoscalar we have $Z\sim N$ such that the up and
down quark terms are suppressed; this is a benefit of the NuSOnG glass
(${\rm SiO}_{2}$) target which is very nearly isoscalar. More specifically,
for ${\rm SiO}_{2}$ we have $Z({\rm O})=8,$ $Z({\rm Si})=14$, $m({\rm O})=15.994$,
$m({\rm Si})=28.0855$. Using $A=Z+N$ we have $(N-Z)/A=(A-2Z)/A$
for the prefactor in Eq.~(\ref{eq:deltaf3a}) which yields $(N-Z)/A\sim-0.000375$
for $O$ and $(N-Z)/A\sim0.00304$ for Si.

In Eq.~(\ref{eq:deltaf3a}) the PDFs $\{ u^{p/A},d^{p/A},...\}$
represent quark distributions bound in a nucleus $A$. With a single
nuclear target, we can determine the CSV term $\delta I^{A}$ for
this specific $A$; measurements on different nuclear targets would
be required in order to obtain the $A$ dependence of $\delta I^{A}$
if we need to scale to a proton or isoscalar target.

Thus, an extraction of any isospin violation $\delta I^{A}$ requires
a careful separation of these contributions from the strange, charm,
and higher order terms. Theoretical NLO calculations for $\Delta xF_{3}$
are available; thus the ${\cal O}\left(\alpha_{S}\right)$ corrections
can be addressed. Additionally, NuSOnG can use the dimuon process
$(\nu N\to\mu^{+}\mu^{-}X)$ to constrain the strange sea.

In conclusion we find that while this is a challenging measurement,
NuSOnG's high statistics measurement of $\Delta xF_{3}$ should provide
a window on CSV which is relatively free of large experimental systematics.
We emphasize that $\Delta xF_{3}$ may be extracted from a single
target, thereby avoiding the complications of introducing nuclear
corrections associated with different targets. This is in contrast
to the other measurements discussed below. However, if we desire to
rescale the $\delta I^{A}$ effects to a different nucleus $A$, then
multiple targets would be required.

\subsection{Measurement of $\Delta F_{2}\equiv\frac{5}{18}\, F_{2}^{CC}(x,Q^{2})-F_{2}^{NC}(x,Q^{2})$}

A separate determination of CSV can be achieved using the measurement
of $F_{2}$ in CC and NC processes via the relation:

\begin{eqnarray}
\Delta F_{2} & \equiv & \frac{5}{18}\, F_{2}^{CC,A}(x,Q^{2})-F_{2}^{NC,A}(x,Q^{2})\nonumber \\
 & \simeq & \frac{1}{6}x\frac{\left(N-Z\right)}{A}\left[\left(u^{p/A}-d^{p/A}\right)+\left(\bar{u}^{p/A}-\bar{d}^{p/A}\right)\right]\nonumber \\
 & + & \frac{1}{6}x\, s^{+,A}(x)-\frac{1}{6}x\, c^{+,A}(x)+\frac{1}{6}x\,\frac{N}{A}\,\delta I^{A}\nonumber \\
 & + & {\cal O}(\alpha_{s})\label{eq:deltaf2}\end{eqnarray}
 with the definitions: \[
F_{2}^{CC,A}=\frac{1}{2}\,\left[F_{2}^{\nu A}+F_{2}^{\bar{\nu}A}\right]\]

\[
F_{2}^{NC,A}=F_{2}^{\ell A}\]
 In Eq.~(\ref{eq:deltaf2}), the first term is proportional to $(N-Z)/A$
which vanishes for an isoscalar target. The second and third terms
are proportional to the heavy quark distributions $s^{+,A}$ and
$c^{+,A}$. The next term is the CSV contribution which is proportional
to $\delta I^{A}$ given in Eq.~(\ref{eq:csv}). It is curious that
this has the same form as the CSV contribution for $\Delta xF_{3}$
of Eq.~(\ref{eq:deltaf3a}). Finally, the last term represents the
higher-order QCD corrections.

While the character of the terms on the LHS of Eq.~(\ref{eq:deltaf3})
and Eq.~(\ref{eq:deltaf2}) are quite similar, the systematics of
measuring $\Delta F_{2}$ may differ substantially from that of $\Delta xF_{3}$.
For example, the measurement of $\Delta F_{2}$ requires the subtraction
of structure functions from two entirely different experiments. The
CC neutrino--nucleon data are extracted from heavy nuclear targets
(to accumulate sufficient statistics); as such, these data are generally
subject to large nuclear corrections so that the heavy targets can
be related to the isoscalar $N=\frac{1}{2}(p+n)$ limit. Conversely,
the NC charged-lepton--nucleon process proceeds via the electromagnetic
interaction. Therefore sufficient statistics can be obtained for light
targets including $H$ and $D$ and no large heavy target corrections
are necessary. Therefore, we must use the appropriate nuclear correction
factors when we combine $F_{2}^{CC}$ and $F_{2}^{NC}$, and this
will introduce a systematic uncertainty.

Separately, the heavy quark production mechanism is different in the
CC and NC processes. Specifically, in the CC case we encounter the
process $s+W^{+}\to c$ where the charm mass threshold kinematics
must be implemented. On the other hand, the NC process is $c+\gamma\to c$
which is proportional to the charm sea distribution and has different
threshold behavior than the CC process. Even though the charm production
process is modeled at NLO, the theoretical uncertainties which this
introduces can dominate precision measurements.

\subsection{Other Measurements of CSV}

We very briefly survey other measurements of CSV in comparison to
the above.

The measurement of the lepton charge asymmetry in $W$ decays from
the Tevatron can constrain the up and down quark distributions \cite{Abe:1998rv,Bodek:1999bb}.
In this case, the extraction of CSV constraints is subtle; while isospin
symmetry is not needed to relate $p$ and $\bar{p}$, this symmetry
is typically used in a global fit of the PDFs to reduce data on heavy targets
to $p$.

In the limit that all the data in the analysis were from proton
targets, CSV would not enter; hence this limit only arises indirectly
from the mix of targets which enter a global fit. At present, while
much of the data does come from proton targets (H1, ZEUS, CDF, D0),
there are some data sets from both $p$ and $d$ (BCDMS, NMC, E866),
and some that use heavier targets (E-605, NuTeV) \cite{Owens:2007kp,Pumplin:2002vw}.
Thus, an outstanding question is if CSV were present, to what extent
would this be {}``absorbed'' into a global fit. The ideal
procedure would be to parameterize the CSV and include this in a
global analysis. While this step has yet to be implemented, there
is a recent effort to include the nuclear corrections as a dynamic
part of a global fit \cite{Schienbein:2007fs}.

Additionally, NMC measures $F_{2}^{n}/F_{2}^{p}$ data which has an
uncertainty of order a few percent \cite{Allasia:1990nt}. There are
also fixed-target Drell-Yan experiments such as NA51 \cite{Baldit:1994jk}
and E866 \cite{Hawker:1998ty} which are sensitive to the ratio $\bar{d}/\bar{u}$
in the range $0.04<x<0.27$. We will soon have LHC data ($pp$) to
add to our collection, thus providing additional constraints in a
new kinematic region.

\subsection{Conclusions on Charge Symmetry Violation}

NuSOnG will be able to provide high statistics DIS measurements across
a wide $x$ range. Because the target material (SiO$_{2}$) is nearly
isoscalar, this will essentially allow a direct extraction of the
isoscalar observables.

$\Delta xF_{3}$ is one of the cleaner measurements of CSV in terms
of associated experimental systematic uncertainties as this measurement
can be extracted from a single target. The challenge here will be
to maximize the event samples.

The measurement of $\Delta F_{2}$ is more complicated as this must
combine measurements from both CC and NC experiments which introduces
nuclear correction factors \cite{Brodsky:2004qa,Schienbein:2007fs}. 
Since NuSOnG will provide high statistics
$F_{2}^{CC}$measurements for a variety of $A$ targets, this will
yield an alternate handle on the CSV and also improve our understanding
of the associated nuclear corrections.

The combination of these measurements, together with external constraints,
will yield important information on this fundamental symmetry.

\section{Measurements of the Heavy Quarks}

\subsection{Measurement of the Strange Sea\label{subsec:strangesea}}

Charged current neutrino-induced charm production, $(\nu/\bar{\nu})N\rightarrow\mu^{+}\mu^{-}X$,
proceeds primarily through the sub-processes $W^{+}s\to c$ and $W^{-}\bar{s}\to\bar{c}$
(respectively), so this provides a unique mechanism to directly probe
the $s(x)$ and $\bar{s}(x)$ distributions. Approximately 10\% of
the time the charmed particles decay into $\mu+X$, adding a second
oppositely signed muon to the CC event's final state. These {}``dimuon''
events are easily distinguishable, and make up approximately 1\% of
the total CC event sample. Hence, the recent high statistics dimuon
measurements \cite{Bazarko:1994tt,Goncharov:2001qe,Tzanov:2003gq,Vilain:1998uw,Astier:2000us}
play an essential role in constraining the strange and anti-strange
components of the proton. On NuSOnG, the dimuon data will be used
in the same manner.

Distinguishing the difference between the $s(x)$ and $\bar{s}(x)$
distributions, \begin{equation}
xs^{-}(x)\equiv xs(x)-x\overline{s}(x),\end{equation}
 is necessary for the PW style analysis. This analysis is sensitive
to the integrated strange sea asymmetry, \begin{equation}
S^{-}\equiv\int_{0}^{1}s^{-}(x)dx,\end{equation}
 through its effect on the denominator of the PW ratio, as has been
recognized in numerous references \cite{Zeller:2002du,Barone:1999yv,McFarland:2003jw,Davidson:2001ji,Olness:2003wz}).

\begin{figure}[t]
\includegraphics[width=0.48\textwidth,keepaspectratio]{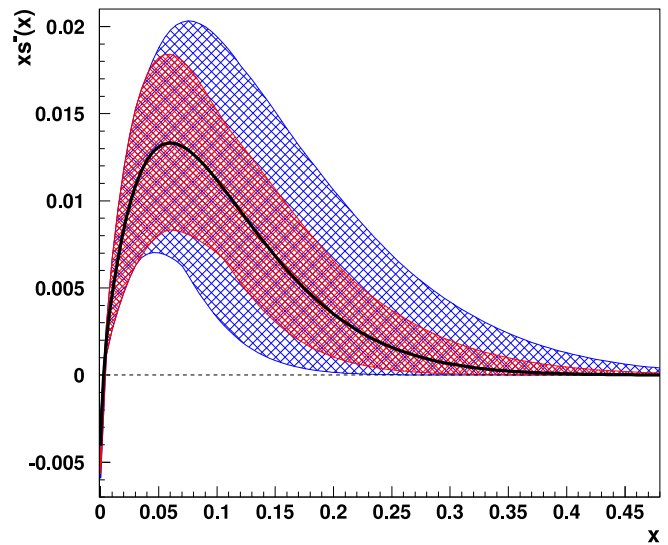}

\caption{\label{sasym} NuTeV measurement of $xs^{-}(x)$ vs $x$ at $Q^{2}=16\, {\rm GeV}^{2}$. 
Outer band is combined errors, inner band is without $B_{c}$
uncertainty.}
\end{figure}

The highest precision study of $s^{-}$ to date is from the NuTeV
experiment \cite{Goncharov:2001qe,Mason:2007zz}. 
The sign selected beam allowed measurement
of the strange and anti-strange seas independently, recording 5163
neutrino-induced dimuons, and 1380 antineutrino induced dimuon events
in its iron target. Figure~\ref{sasym} shows the fit for asymmetry
between the strange and anti-strange seas in the NuTeV data.

%

The integrated strange sea asymmetry from NuTeV has a positive central
value: $0.00196\pm0.00046$ (stat) $\pm0.00045$ (syst) $_{-0.00107}^{+0.00148}$
(external). In NuSOnG, as in NuTeV, the statistical error will be
dominated by the antineutrino data set and is expected to be about
0.0002. The systematic error is dominated by the $\pi$ and $K$ decay-in-flight
subtraction. This can be addressed in NuSOnG through test-beam measurements
which will allow a more accurate modeling of this background, as well
as applying the techniques of CCFR to constrain this rate \cite{Sandler:1990ck,Sandler:1992wj,Sandler:1992ai}.
We expect to be able to reduce this error to about 0.0002. The combination
of these reduces the total error by about 10\%, because the main contributon
comes from the external inputs.

The external error on the measurement is dominated by the error on
the average charm semi-muonic branching ratio, $B_{c}$: \begin{equation}
B_{c}=\Sigma_{i}\int\phi(E)f_{i}(E)B_{\mu-i}dE,\end{equation}
 where $\phi$ is the neutrino flux in energy bins, $f_{i}$ is the
energy dependent production fraction for each hadron, and $B_{\mu-i}$
is the semi-muonic branching ratio for each hadron. In the NuTeV analysis,
this is an external input, with an error of about 10\%. To make further
progress, this error must be reduced.

\begin{figure}[t]
\includegraphics[clip,width=0.48\textwidth,keepaspectratio]{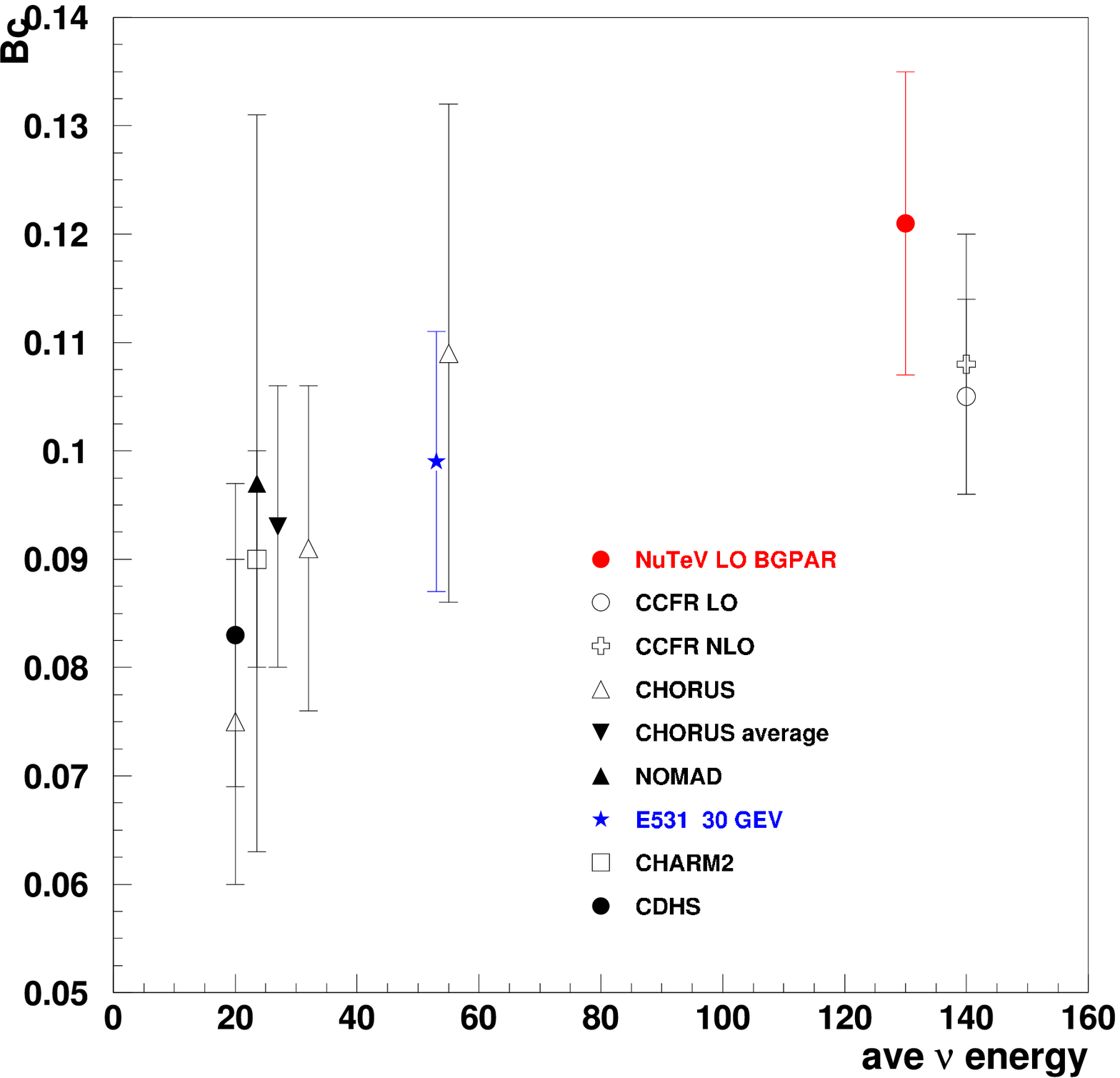}

\caption{\label{Bc} 
World measurements of $B_{c}$. 
See refs.~\cite{Bazarko:1994tt,Rabinowitz:1993xx, Abramowicz:1982zr, Bolton:1997pq,Astier:2000us, Vilain:1998uw, KayisTopaksu:2002xq}.}
\end{figure}

Fig.~\ref{Bc} shows the world measurements of $B_{c}$, taken from
references \cite{Bazarko:1994tt,Rabinowitz:1993xx, Abramowicz:1982zr, Bolton:1997pq,Astier:2000us, Vilain:1998uw, KayisTopaksu:2002xq}.  Measuring $B_{c}$
directly requires the capability to resolve the individual charmed
particles created in the interaction. The best direct measurements
are from emulsion. This kind of measurement has been performed in
past experiments (E531, Chorus) using emulsion detectors \cite{Bolton:1997pq,KayisTopaksu:2002xq},
where the decay of the charmed meson is well tagged. Since the cross
section for charmed meson production is energy dependent, it is important
to make a measurement near the energy range of interest. The NuTeV
strange sea asymmetry study used a re-analysis of 125 charm events
measured by the FNAL E531 experiment \cite{Bolton:1997pq} in the energy
range of the NuTeV analysis ($E_{\nu}>20$ GeV). $B_{c}$ has also
been constrained through indirect measurement via fits.

For NuSOnG, our goal is to reduce the error on $B_{c}$ using an \textit{in
situ} measurement on glass by at least a factor of 1.5. One method
is to include $B_{c}$ as a fit parameter in the analysis of the dimuon
data. The unprecedentedly high statistics will allow a fit as a function
of neutrino energy for the first time. Dimuons from high $x$ neutrino
DIS almost exclusively result from scattering off valence quarks,
such that the dimuon cross section in that region isolates $B_{c}$
from the strange sea. In dimuon fits, the assumption is then taken
that $B_{c-\nu}=B_{c-\overline{\nu}}$, $B_{c}$ may be measured directly
from the dimuon data.

Unfortunately, antineutrino charm production is not well measured
by past experiments. This leads to concerns about the assumption that
$B_{c-\nu}=B_{c-\overline{\nu}}$. An example of a potential source
of difference in neutrino and antineutrino mode, consider that $\nu n\rightarrow\mu^{-}\Lambda_{c}$
has no analogous reaction in the antineutrino channel.

These arguments provide the motivation for including a high resolution
target/tracker in the NuSOnG design that can directly measure the
semileptonic branching ratio to charm in both $\nu$ and $\bar{\nu}$
running modes. There are two feasible detector technologies. The first
is to use emulsion, as in past experiments. This is proven technology
and scanning could be done at the facility in Nagoya, Japan. The second
is to use the NOMAD-STAR detector \cite{Barichello:2003gu,Ellis:2003vq}
or a similar
detector. This is a 45 kg silicon vertex detector which ran in front
of the NOMAD experiment. The target was boron carbide interleaved
with the silicon. This detector successfully measured 45 charm events
in that beam, identifying $D^{+}$, $D^{0}$ and $D_{s}$. A similar
detector of this size in the NuSOnG beam would yield about 900 $\nu$
events and 300 $\bar{\nu}$ events. This has the advantage of being
a low-Z material which is isoscalar and close in mass to the SiO$_{2}$
of the detector.

%

\subsection{Strange Quark Contribution to the Proton Spin}

An investigation of the strange quark contribution to the elastic
vector and axial form factors of the proton is possible in NuSOnG, by
observing NC elastic and CC quasi-elastic scattering events; namely
$\nu p \rightarrow \nu p$ and $\nu n \rightarrow \mu^- p$ events in
neutrino mode, and $\bar{\nu} p \rightarrow \bar{\nu} p$ and
$\bar{\nu} p \rightarrow \mu^+ n$ events in antineutrino mode.  The
motivation for making this measurement comes from a number of recent
(and not so recent) studies in proton structure.

Over the last 15 years a 
tremendous effort has been made at MIT-Bates, Jefferson Lab, and Mainz
to measure the strange quark contribution to the vector form factors
(that is, the electromagnetic form factors) of the proton via
parity-violating electron scattering from protons, deuterons, and
$^4$He~\cite{Mueller:1997mt,Hasty:2001ep,Spayde:2003nr,Ito:2003mr,Aniol:2004hp,Maas:2004ta,Maas:2004dh,Armstrong:2005hs,Acha:2006my,Aniol:2005zg,Aniol:2005zf}.
The technique is to observe the parity-violating beam spin asymmetry
in elastic scattering of longitudinally polarized electrons from these
unpolarized targets; this asymmetry is caused by an interference
between the one-photon and one-$Z$ exchange
amplitudes~\cite{Musolf:1993tb}.  As a result, the weak
neutral current analog of the electromagnetic form factors of the
proton may be measured and this gives access to the strange quark
contribution.  This worldwide experimental program will soon be
complete.  The results available to date (from global
analyses~\cite{Pate:2008va,Liu:2007yi,Young:2006jc}) indicate a small
(and nearly zero) contribution of the strange quarks to the elastic
electric form factor, $G_E^s$; this is not surprising, as the total
electric charge in the proton due to strange quarks is zero.  At the
same time, these same data point to a small but likely positive
contribution of the strange quarks to the elastic magnetic form
factor, $G_M^s$, indicating a small positive contribution of the
strange quarks to the proton magnetic moment.  Due to the prominent
role played by the $Z$-exchange amplitude, these experiments are also
sensitive to the strange quark contribution to the elastic {\em axial}
form factor, which is related to the proton spin structure.

It is now well established by 
leptonic deep inelastic scattering experiments that the spins of the
valence and sea quarks in the proton together contribute about 30\% of
the total proton intrinsic angular momentum of $\hbar/2$.  The strange
quark contribution is estimated to be about -10\% in inclusive DIS (an
analysis which makes use of SU(3)-flavor
symmetry)~\cite{Leader:2005ci}, but is found to be approximately zero
in semi-inclusive DIS (an alternative analysis which makes no use of
SU(3) but needs fragmentation functions
instead)~\cite{Airapetian:2008qf}.  A recent global
analysis~\cite{deFlorian:2008mr} which made use of both inclusive and
semi-inclusive DIS and which allowed for the possibility of
SU(3)-flavor violation found no need in the data for any violation of
SU(3) and indicated a small negative contribution of strange quarks to
the proton spin.  In the deep inelastic context, the contribution
strange quarks make to the proton spin is encapsulated in the
helicity-difference strange quark parton distribution function,
$$\Delta s(x) = s^\rightarrow (x) - s^\leftarrow (x)$$ where
$s^\rightarrow (x)$ [$s^\leftarrow (x)$] is the probability density
for finding a strange quark of momentum fraction $x$ with its spin
parallel [anti-parallel] to the proton spin.  The axial current
relates the first moment of this parton distribution function to the
value of the strange quark contribution to the elastic axial form
factor of the proton~\cite{Anselmino:1994gn}, $G_A^s$, at $Q^2=0$:
$$\int_0^1 dx \Delta s(x) = G_A^s(Q^2=0).$$ The strange quark
contribution to the elastic axial form factor can be measured by
combining data from neutrino NC elastic scattering from the proton
with data from parity-violating elastic $\vec{e}p$
scattering~\cite{Pate:2003rk}.  In this way the strange
quark spin contribution to the proton spin can be measured in a
completely independent way using low-$Q^2$ elastic scattering instead
of high-$Q^2$ deep inelastic scattering.  An analysis done using
this method~\cite{Pate:2008va} indicates that $G_A^s$ may in
fact be negative at $Q^2=0$ but this conclusion is not definitive
due to the limitations of the currently available neutrino data.

Since the neutrino experiments will undoubtedly be carried out on
nuclear targets (perhaps carbon or argon), then the extraction of the
properties of the proton from these data needs to be done with care.
Recent theoretical investigations point to the idea of measuring the
ratio of NC to CC yields; nuclear effects appear to largely cancel in
this ratio~\cite{Jachowicz:2007ek}, leaving behind the ratio that
would have been obtained on nucleon targets.

The only available data on neutrino NC elastic scattering is from the
BNL E734 experiment~\cite{Ahrens:1986xe}; the uncertainties reported
from that experiment are considerable and limit the preciseness of any
extraction of $G_A^s$ based on them.  If NuSOnG can provide more
precise measurements of NC elastic scattering extended to lower $Q^2$
then the promise of this analysis technique can be fulfilled.


\subsection{Measurements of the Charm Sea: }

\subsubsection{Charm Production}

We can also study the charm sea component of the proton which can
arise from the gluon splitting process $g\to c\bar{c}$ producing
charm constituents inside the 
proton.\cite{Conrad:1997ne,Mishra:1989rr,Alton:2001fu}
In a measurement complementary to the above strange sea extraction,
the charm sea, $c(x,\mu)$, can be measured using the following process:\[
\begin{array}{rl}
\nu_{\mu}+c\to\nu_{\mu}+ & c\\
 & \hookrightarrow s+\mu^{+}+\nu_{\mu}\end{array}\ . \]
 In this process, we excite a constituent charm quark in the proton
via the NC exchange of a $Z$ boson; the final state charm quark then
decays semi-leptonically into $s\mu^{+}\nu_{\mu}$. We refer to this
process as Wrong Sign Muon (WSM) production as the observed muon is
typically the opposite sign from the expected $\nu_{\mu}d\to\mu^{-}u$
DIS process. For antineutrino beams, there is a complementary process
$\bar{\nu}_{\mu}+\bar{c}\to\bar{\nu}_{\mu}+\bar{c}$ with a subsequent
$\bar{c}\to\bar{s}+\mu^{-}+\bar{\nu}_{\mu}$ decay with yields a WSM
with respect to the conventional $\bar{\nu}_{\mu}u\to\mu^{+}d$ process.
Here, the ability of NuSOnG to have sign-selected beams is crucial
to this measurement as it allows us  to distinguish 
the secondary muons, and thus extract the charm-sea component.

In the conventional implementation of the heavy quark PDFs, the charm
quark becomes an active parton in the proton when the scale $\mu$
is greater than the charm mass $m_{c}$; \emph{i.e.} $f_{c}(x,\mu)$
is nonzero for $\mu>m_{c}$. Additionally, we must {}``rescale''
the Bjorken $x$ variable as we have a massive charm in the final
state. The original rescaling procedure is to make the substitution
$x\to x(1+m_{c}^{2}/Q^{2})$ which provides a kinematic penalty for
producing the heavy charm quark in the final state.\cite{Barnett:1976ak}
As the charm is pair-produced by the $g\to c\, \bar{c}$ process, there
are actually two charm quarks in the final state---one which is observed
in the semi-leptonic decay, and one which 
decays hadronically and is part of the hadronic shower.  
Thus, the appropriate rescaling is not $x\to x(1+m_{c}^{2}/Q^{2})$
but instead $x\to\chi=x(1+4m_{c}^{2}/Q^{2})$; this rescaling is implemented
in the ACOT--$\chi$ scheme, for example.\cite{Amundson:1998zk,Amundson:2000vg,Tung:2001mv}
The factor $(1+4m_{c}^{2}/Q^{2})$ represents a kinematic suppression
factor which will suppress the charm process relative to the lighter
quarks.

The differential cross section for NC neutrino scattering is \begin{eqnarray*}
\frac{d\sigma}{d\xi dy}(\nu p & \to & \nu c)=\frac{G_{F}^{^{2}}M_{N}E_{\nu}}{\pi}\, R_{Z}^{2}(Q^{2})\,\times\\
 & \times & \left[g_{L}^{2}+g_{R}^{2}(1-y)^{2}-\frac{1}{2}\left(2g_{L}g_{R}\right)\frac{M_{N}}{E_{\nu}}\right]\,\xi\, c(\xi,\mu),\end{eqnarray*}
 where $g_{L}=t_{3}-Q_{c}^{2}\sin^{2}\theta_{W}$, $g_{R}=-Q_{c}^{2}\sin^{2}\theta_{W}$,
and for charm $t_{3}=1/2$ and $Q_{c}=2/3$. The factor $R_{Z}(Q^{2})=1/(1+Q^{2}/M_{Z}^{2})$
arises from the $Z$-boson propagator. The corresponding result for
the anti-charm is given with the substitutions $g_{L}\leftrightarrow g_{R}$
and $c\leftrightarrow\bar{c}$. 

In the limit we can neglect the $M_{N}/E_{\nu}$ term we have the
approximate expressions for the total cross section:\cite{Mishra:1989rr}
\begin{equation}
\sigma(\nu p\to\nu c)\sim\frac{G_{F}^{2}M_{N}E_{\nu}}{\pi}\,(0.129)\, C\label{eq:sigCharm}\end{equation}
 and

\begin{equation}
\sigma(\nu p\to\nu\bar{c})\sim\frac{G_{F}^{2}M_{N}E_{\nu}}{\pi}\,(0.063)\,\bar{C}\label{eq:sigAntiCharm}\end{equation}
 with $C=\int_{\xi_{min}}^{1}\xi\, c(\xi,\mu)\, d\xi$ and $\bar{C}=\int_{\xi_{min}}^{1}\xi\,\bar{c}(\xi,\mu)\, d\xi$.
We take $\xi=x(1+4m_{c}^{2}/Q^{2})$ and $\xi_{min}=m_{c}^{2}/(2M_{N}\nu)$. 

We will be searching for the WSM signal compared to the conventional
charged-current DIS process; therefore it is useful to benchmark the
rate for WSM production by comparing this to the the usual charged-current
DIS process, 

\begin{eqnarray}
\frac{d\sigma}{dx\, dy}(\nu p & \to & \mu^{-}X)=\frac{G_{F}^{^{2}}M_{N}E_{\nu}}{\pi}\, R_{W}^{2}(Q^{2})\,\times\nonumber \\
 & \times & \left[q(x)+(1-y)^{2}\bar{q}(x)\right]\label{eq:sigDIS}\end{eqnarray}
with $R_{W}(Q^{2})=1/(1+Q^{2}/M_{W}^{2})$. We can again integrate
over $x$ and $y$ to obtain an estimate of the total cross section
in terms of the integrated PDFs as in Eq.~(\ref{eq:sigCharm}) and
Eq.~(\ref{eq:sigAntiCharm}): 

\begin{eqnarray}
\sigma(\nu N & \to & \mu^{-}X)\sim\frac{G_{F}^{^{2}}M_{N}E_{\nu}}{\pi}\, R_{W}^{2}(Q^{2})\,\times\nonumber \\
 & \times & \frac{1}{2}\left[U+D+2S+\frac{1}{3}(\bar{U}+\bar{D}+2\bar{C})\right]\label{eq:sigDIS2}\end{eqnarray}
where $\{U,D,S\}$ are defined analogously to $C$, and we have used
$N=\frac{1}{2}(p+n)$ for an isoscalar target. 

The relative rate for NC charm production is determined by the above
factors together with a ratio of integrated PDFs. For a mean neutrino
energy of 100~GeV, the massive charm cross section is down a factor
of $\sim0.005$ compared to the total inclusive cross section. As
the muon from the NC charm process is a secondary muon, we must additionally
fold in the semi-leptonic branching ratio $B_{c}\sim10\%$, and the
acceptance factor of observing the secondary muon in the detector
($A_{\mu}$$\sim$20\%).\cite{Bazarko:1994tt} Combining the relevant
factors, we estimate the rate for NC charm production is approximately
a factor of $10^{-4}$ as compared to the CC DIS process. Thus, for
an anticipated design of 600M $\nu_{\mu}$ CC events, one would expect
on the order of 60K NC charm events. 
This estimate is also consistent with a direct scaling from the 
NuTeV result of Ref.~\cite{Alton:2001fu}.

\subsubsection{Backgrounds}

Extrapolating from  investigations by CCFR~\cite{Mishra:1989rr}, 
and NuTeV~\cite{Alton:2001fu}, 
the dominant background for the measurement of the charm sea 
comes from $\bar{\nu}_{\mu}$ contamination. In these
studies, it was determined that by demanding $E_{vis}>100$~GeV, the
background rate could be reduced to $2.3\times10^{-4}$. Other background
processes include $\nu_{e}$ induced dilepton production, mis-identified
dimuon events, and NC interactions with a $\pi/K$ decay in the hadron
shower; these processes contribute approximately an additional $1.5\times10^{-4}$
to the background rate. As compared to CCFR and NuTeV, the NuSOnG design has
a number of improvements such as lower mass density for improved shower
measurement; hence, comparable background reductions should be achievable.

\subsubsection{Intrinsic Charm}

\begin{figure}[t]
 \includegraphics[angle=0,width=0.48\textwidth,keepaspectratio]{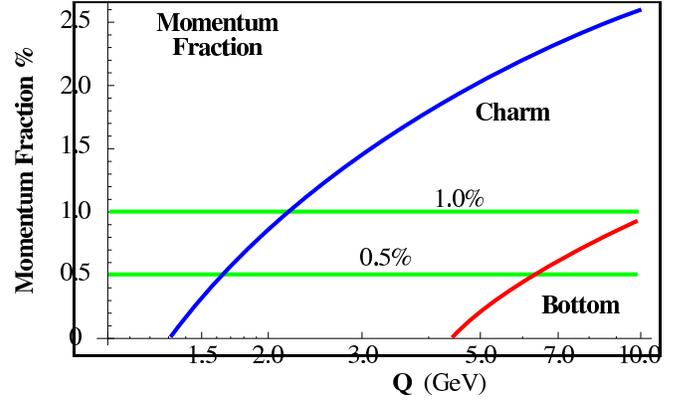}

\caption{Integrated momentum fractions $\int_{0}^{1}x\, f_{i}(x,Q)$ of charm (upper curve)
and bottom (lower curve) PDFs (in percent) vs. $Q$ in GeV. 
Both the quark and antiquark contributions are included. 
 Horizontal lines at 0.5\% and 1.0\% are indicated as this is the 
typical size of postulated intrinsic contributions. 
\label{fig:momFrac2}}

\end{figure}

In the above discussion we have assumed that the charm component of
the proton arises perturbatively from gluons splitting into charm
quark pairs, $g\to c\bar{c}$; in this scenario the charm PDF typically
vanishes at scales below the charm mass ($f_{c}(x,\mu<m_{c})=0$), 
and for $\mu>m_{c}$ all the charm partons arise from gluon splitting.

There is an alternative picture where the charm quarks are taken to
be intrinsic to the proton; in this case there are intrinsic charm
partons present at scales $\mu<m_{c}$. For $\mu>m_{c}$, the charm
PDF is then a combination of this {}``intrinsic'' PDF and the {}``extrinsic''
PDF component arising from the $g\to c\bar{c}$ process.

A number of analyses have searched for an intrinsic charm component
of the proton, and this intrinsic component is typically constrained
to have an integrated momentum fraction less than a percent or two
\cite{Harris:1995jx,Pumplin:2007wg}.

In Figure~\ref{fig:momFrac2} we display the integrated momentum fraction,
$\int_{0}^{1}x\, f_{i}(x,\mu)$, for charm and bottom as a function
of $\mu$ due to the {}``extrinsic'' PDF component arising from
the $g\to c\bar{c}$ or $g\to b\bar{b}$ process. These momentum fractions
start from zero at the corresponding quark mass, and increase slowly
as the partonic components pick up momentum from the gluon splitting
process.

If we are searching for an additional intrinsic component with a momentum
fraction of $\sim1\%$, we will be most sensitive to such a component
in the threshold region where the {}``intrinsic'' component is not
overwhelmed by the {}``extrinsic'' contribution. In this regard,
NuSOnG is well suited to search for these intrinsic terms as it will
provide good statistics in the threshold region. Measuring the charm
production process described above, NuSOnG can attempt to extract
the charm PDF as a function of the $\mu$ scale, and then evolve back
to $\mu=m_{c}$. Three outcomes are possible:

\begin{enumerate}
\item $f_{c}(x,\mu=m_{c})<0$, which would imply the data are inconsistent
with the normal QCD evolution.%
\footnote{If we work at NLO, $f_{c}(x,\mu=m_{c})$ should be strictly greater
than or equal to zero; at NNLO and beyond the boundary conditions
yield a negative PDF of order $\sim\alpha_{s}^{2}$ %
} 
\item $f_{c}(x,\mu=m_{c})=0$, which would imply the data is consistent
with no intrinsic charm PDF. 
\item $f_{c}(x,\mu=m_{c})>0$, which would imply the data is inconsistent
with an intrinsic charm PDF. 
\end{enumerate}
By making accurate measurements of charm induced processes in the
threshold region, NuSOnG can provide a discriminating test to determine
which of the above possibilities is favored. Hence, the high statistics
of NuSOnG in the threshold region are well suited to further constrain
the question of an intrinsic charm component.

\section{Summary and Conclusions}

The NuSOnG experiment can search for ``new physics'' from the keV
through TeV energy scales. This article has focused mainly on the
QCD physics which can be accessed with this new high energy, high
statistics neutrino scattering experiment. During its five-year data acquisition
period, the NuSOnG experiment could record almost one hundred thousand
neutrino-electron elastic scatters and hundreds of millions of deep
inelastic scattering events, exceeding the current world data sample
by more than an order of magnitude.

With this wealth of data, NuSOnG can address a wide variety of topics 
including the following.

\begin{itemize}

\item NuSOnG can increase the statistics of the Elastic Scattering (ES) and
Deeply Inelastic Scattering (DIS) data sets by nearly two orders of magnitude.

\item The unprecedented statistics of NuSOnG  
allow the possibility to perform separate extractions of the 
 structure functions:
$\{  
F_2^{\nu}, xF_e^{\nu}, R_L^{\nu},
F_2^{\bar\nu}, xF_e^{\bar\nu}, R_L^{\bar\nu}
\}$.
This allows us to test many of the symmetries and assumptions which were employed in 
previous structure function determinations.

\item 
NuSOnG will help us to disentangle the nuclear effects which are present in the 
PDFs. Furthermore, this may help us address the long-standing 
tensions between the NC charged-lepton and CC neutrino DIS measurements.

\item High precision NuSOnG  measurements are sensitive to Charge Symmetry Violation (CSV) 
 and other ``new physics'' processes. Such effects can significantly  influence
precision Standard Model  parameter extractions such as $\sin\theta_{W}$.
In particular, $\Delta xF_{3}$ is a sensitive probe of both the heavy quark components,
and CSV effects.

\item NuSOnG dimuon production provides an exceptional probe of the strange
quark PDFs, and the sign-selected beam can separately study $s(x)$ and  $\bar{s}(x)$.
Additionally, NuSOnG can probe the $s$-quark contribution to the proton spin.

\item 
The high statistics of NuSOnG may allow the measurement of the charm sea and an 
method to prove the intrinsic-charm content of the proton. 
While this is a difficult measurement,  the NuSOnG kinematics 
allow the measurement of charm-induced processes in the 
threshold region where the  ``intrinsic'' character can most easily be discerned. 

\end{itemize}

While the above list presents a very compelling physics case for NuSOnG, this is only
a subset of the full range of investigations that can be addressed with this facility.

\subsection*{Acknowledgments}

We thank the following people for their 
informative discussions regarding  neutrino-nucleus interactions.
and their
thoughtful comments on the
development of this physics case: 
Andrei Kataev,
Sergey Kulagin,
P. Langacker,
Roberto Petti,
M.~Shaposhnikov, 
F. Vannucci, 
  and 
J. Wells. 
We acknowledge the support of the following funding agencies for the
authors of this paper: Deutsche Forschungsgemeinschaft, The Kavli
Institute for Theoretical Physics, The United States Department of
Energy, The United States National Science Foundation.


\bibliographystyle{hunsrt}

\bibliography{nusong}{}


\end{document}